\documentclass[a4paper,twocolumn,11pt,unpublished]{quantumarticle}
\pdfoutput=1
\usepackage[utf8]{inputenc}
\usepackage[english]{babel}
\usepackage[T1]{fontenc}
\usepackage{amsmath}
\usepackage{hyperref}
\usepackage{amssymb}

\usepackage[numbers]{natbib}

\usepackage{graphicx}%
\usepackage{dcolumn}%
\usepackage{bm}%
\usepackage{physics}
\usepackage{hyperref}%
\usepackage{color}

\usepackage{ulem} %
\usepackage{multirow}
\usepackage{booktabs}
\usepackage{blkarray}
\usepackage{tabularray}

\usepackage[ruled]{algorithm2e} 
\usepackage{ifthen}

\usepackage[bb=dsserif]{mathalpha}
\usepackage{setspace}

\newcommand{\Ical}{\mathcal{I}}
\newcommand{\Jcal}{\mathcal{J}}
\newcommand{\Fcal}{\mathcal{F}}
\newcommand{\ncnot}{n_{\text{CNOT}}}
\newcommand{\dcnot}{d_{\text{CNOT}}}
\newcommand{\nswpEV}{n_{\text{swp,EV}}}
\newcommand{\niterR}{n_{\text{iter,R}}}
\newcommand{\SvN}{\mathcal{S}_{\text{vN}}}

\begin{document}

\title{Scalable Preparation of Matrix Product States with Sequential and Brick Wall Quantum Circuits}

\author{Tomasz Szo\l{}dra}
\affiliation{Zentrum für Optische Quantentechnologien, Universität Hamburg, Luruper Chaussee 149, 22761 Hamburg, Germany}
\orcid{0000-0002-2897-0506}
\email{tomasz.szoldra@uni-hamburg.de}
\author{Rick Mukherjee}
\affiliation{Department of Physics and Astronomy, University of Tennessee, Chattanooga, TN 37403, USA}
\affiliation{UTC Quantum Center, University of Tennessee, Chattanooga, TN 37403, USA}
\orcid{0000-0001-9267-4421}
\author{Peter Schmelcher}
\affiliation{Zentrum für Optische Quantentechnologien, Universität Hamburg, Luruper Chaussee 149, 22761 Hamburg, Germany}
\affiliation{The Hamburg Centre for Ultrafast Imaging, Universität Hamburg, Luruper Chaussee 149, 22761 Hamburg, Germany}
\orcid{0000-0002-2637-0937}

\begin{abstract}
	Preparing arbitrary quantum states requires exponential resources. Matrix Product States (MPS) admit more efficient constructions, particularly when accuracy is traded for circuit complexity. Existing approaches to MPS preparation mostly rely on heuristic circuits that are deterministic but quickly saturate in accuracy, or on variational optimization methods that reach high fidelities but scale poorly. This work introduces an end-to-end MPS preparation framework that combines the strengths of both strategies within a single pipeline. Heuristic staircase-like and brick wall disentangler circuits provide warm-start initializations for variational optimization, enabling high-fidelity state preparation for large systems. Target MPSs are either specified as physical quantum states or constructed from classical datasets via amplitude encoding, using step-by-step singular value decompositions or tensor cross interpolation. The framework incorporates entanglement-based qubit reordering, reformulated as a quadratic assignment problem, and low-level optimizations that reduce depths by up to $50\%$ and CNOT counts by $33\%$. We evaluate the full pipeline on datasets of varying complexity across systems of 19–50 qubits and identify trade-offs between fidelity, gate count, and circuit depth. Optimized brick wall circuits typically achieve the lowest depths, while the optimized staircase-like circuits minimize gate counts. Overall, our results provide principled and scalable protocols for preparing MPSs as quantum circuits, supporting utility-scale applications on near-term quantum devices.
\end{abstract}

\maketitle

\section{Introduction}
\label{sec:introduction}

\begin{figure*}
	\includegraphics[width=\linewidth]{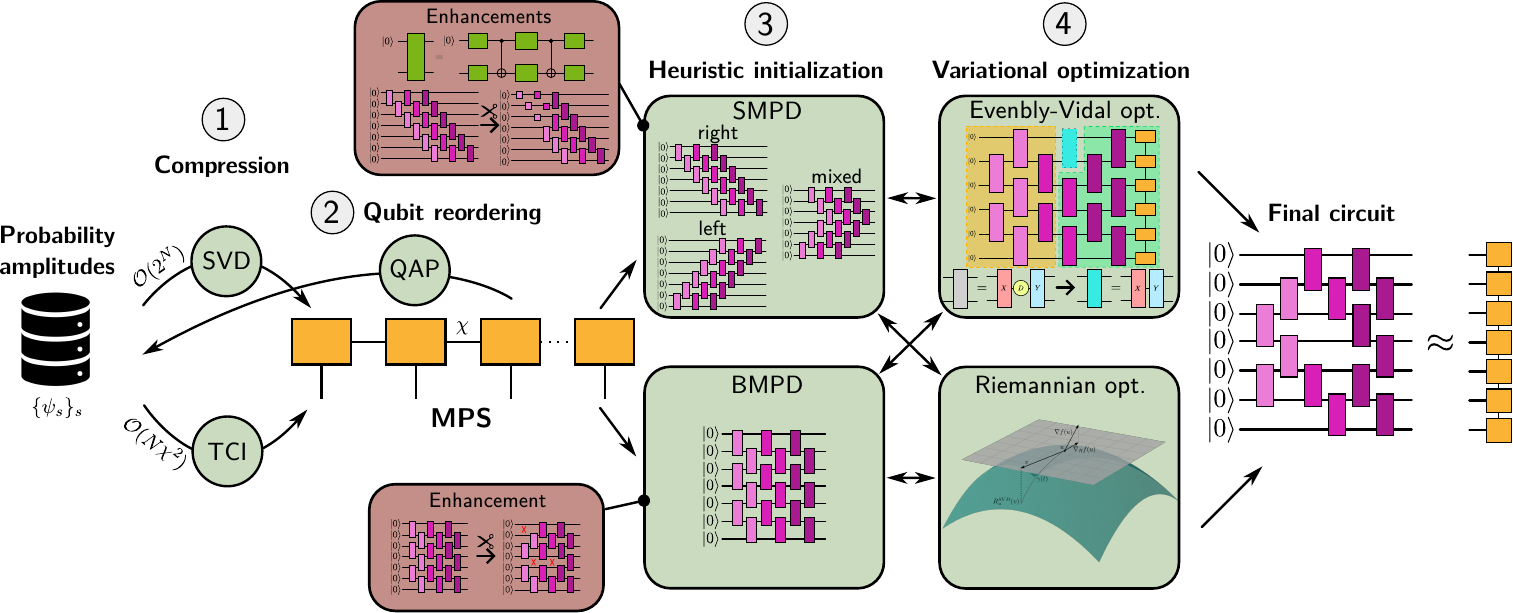}
\caption{Quantum state preparation pipeline. (1) $\mathcal{O}(2^N)$ input probability amplitudes are compressed into an MPS using truncated singular value decomposition (SVD) or tensor cross interpolation (TCI), the latter requiring only $\mathcal{O}(N\chi^2)$ values.
	(2) Qubits are reordered to minimize quantum mutual information between distant pairs; the optimal permutation solves a quadratic assignment problem (QAP) and defines a new MPS.
	(3) Heuristic circuits are built layer-by-layer from the MPS, giving either sequential (SMPD) or brick wall (BMPD) architectures. For SMPD, left-, right-, and mixed-canonical MPS forms yield different gate layouts. "Enhancements" of SMPD include isometric gate decompositions using 2 CNOTs (instead of 3 for a generic two-qubit unitary) and removal of gates not affecting the fidelity, reducing CNOT depth and count; the latter also applies to BMPD.
	(4) Fidelity with the target MPS is variationally optimized, initialized from the heuristic circuits, using the Evenbly-Vidal or Riemannian methods. The optimized circuit may be optionally fed back to the heuristic method to generate a new layer and continue with the optimization. The final circuit approximates the MPS.
}
	\label{fig:full_pipeline_scheme}
\end{figure*}

Efficient quantum state preparation (\textbf{QSP}) is essential for many applications of quantum algorithms with a potential quantum advantage over classical methods, including the Grover's search algorithm \cite{grover_fast_1996, nielsen_quantum_2010}, quantum computational chemistry \cite{aspuruguzik_simulated_2005, mcardle_quantum_2020}, solving linear systems \cite{harrow_quantum_2009,yalovetzky_solving_2024}, optimization \cite{farhi_quantum_2014,koch_resource_2025}, and quantum machine learning \cite{schuld_introduction_2014,peralgarcia_systematic_2024}. On the other hand, it is known that preparation of an arbitrary quantum state requires resources scaling exponentially with the number of qubits \cite{shende_synthesis_2006,plesch_quantum_2011}, which in some cases can even dominate the overall cost of the algorithm execution.

Matrix Product States (\textbf{MPS}) \cite{fannes_finitely_1992, schollwoeck_densitymatrix_2011} offer a way around the exponential computational bottleneck for physically relevant classes of states with limited (area-law) entanglement \cite{hastings_area_2007}, such as groundstates of one-dimensional local gapped Hamiltonians \cite{verstraete_matrix_2006}, requiring storage of $\mathcal{O}(N\chi^2)$ elements, where $N$ is the number of qubits and $\chi$ is a finite bond dimension. From a classical standpoint, smooth functions possess a low-rank MPS representation when encoded as quantum amplitudes \cite{ritter_quantics_2024}, with examples including $\sin(x)$ represented by $\chi=2$ MPS \cite{khoromskij_quantics_2011} and $d$-th order polynomials exactly mapping to $\chi=d+1$ MPS \cite{grasedyck_polynomial_2010, oseledets_constructive_2010}. The class of MPS is particularly useful in the context of QSP - any MPS can be exactly implemented by a quantum circuit of depth $\mathcal{O}(4N\chi^2)$ \cite{schoen_sequential_2005,green_quantum_2025}. Note that QSP based on Tree Tensor Networks (\textbf{TTN}), a more expressive generalization of MPS to tensor networks without closed loops, is also an active area of research \cite{sugawara_embedding_2025,manabe_state_2025,ballarin_efficient_2025}. Alternative approaches, such as Tucker Iterative Quantum State Preparation (Q-Tucker) \cite{blank_system_2025,blank_quantum_2026}, likewise appear very promising for more general classes of states than MPS.

The presence of noise in real quantum hardware \cite{cai_quantum_2023} motivates the need for even shallower circuits and lower gate counts. Addressing this issue, several approximate schemes of MPS preparation have been developed \cite{ran_encoding_2020,araujo_low-rank_2024,bendov_approximate_2024,rudolph_decomposition_2023,melnikov_quantum_2023,malz_preparation_2024,mansuroglu_preparation_2025,wei_state_2025,jaderberg_variational_2025} (see also \cite{smith_crossing_2022,zhang_qubit_2022,wall_quantum_2024,anselme_combining_2024,bohun_scalable_2025,scheer_renormalization_2025} for experiments), and here we focus on two promising categories of such methods. (i) Heuristic schemes that aim to "distentangle" the MPS by applying gates that systematically drive it closer to a product state, referred to as Matrix Product Disentanglers (\textbf{MPD}). Inverse (hermitian conjugate) of such operations, applied to the initial product state of $\ket{0}^{\otimes N}$, yields an approximate state preparation circuit with e.g. a staircase-like sequential (\textbf{SMPD}) \cite{ran_encoding_2020} or brick wall (\textbf{BMPD}) \cite{mansuroglu_preparation_2025} architecture of nearest-neighbor gates. While deterministic and classically simulable, these constructions typically suffer from a quick saturation of fidelity with an increasing number of elementary layers. (ii) Circuits whose fidelity with the target MPS is variationally maximized, employing techniques such as the Evenbly-Vidal (\textbf{EV}) \cite{evenbly_algorithms_2009, shirakawa_automatic_2024} and the Riemannian (\textbf{R}) \cite{luchnikov_qgopt_2021, melnikov_quantum_2023} optimization on unitary manifolds. Although these schemes in principle permit arbitrarily high fidelities (provided the parametrized circuit is expressible enough), with random initialization they are affected by a fast growth of computational complexity with the system size, and vanishing gradients/barren plateaus in the loss landscape \cite{mcclean_barren_2018}, leading to severe problems with trainability and scalability. 

It has been shown that initializing the optimization from a pre-optimized circuit obtained with a heuristic method, and subsequently only refining the approximate solution, is a reliable practical way to mitigate barren plateaus and prepare the MPS \cite{rudolph_decomposition_2023}. However, to our knowledge, so far only the SMPD circuits and the EV optimization algorithm have been combined in such a way \cite{bendov_approximate_2024, rudolph_decomposition_2023, iaconis_tensor_2023, jumade_data_2023, green_quantum_2025}. This work aims to fill the existing gaps by introducing and evaluating combinations of the sequential circuits with Riemannian optimizers (potentially increasing fidelity), as well as applying both optimization techniques also to the brick wall heuristic circuits offering better circuit depths. To that end, we introduce a QSP pipeline presented in Fig.~\ref{fig:full_pipeline_scheme}.

\subsection{QSP pipeline overview}
\label{subsec:qsppipelineoverview}

The protocol in Fig.~\ref{fig:full_pipeline_scheme} starts either from a known target MPS, such as a physical groundstate from DMRG \cite{white_density_1992, white_density_1993}, or by constructing an MPS from given probability amplitudes in Step~1, possibly from a classical dataset. This can be done by (a) reading all $2^N$ amplitudes and applying successive singular value decompositions (\textbf{SVD}) at $\mathcal{O}(N\chi^3)$ cost \cite{oseledets_tensor_2011, schollwoeck_densitymatrix_2011}, or (b) using tensor cross interpolation (\textbf{TCI})  \cite{oseledets_tt-cross_2010} to build the MPS with only $\mathcal{O}(N\chi^2)$ probability amplitude queries.

In Step 2, the qubits are permuted based on pairwise quantum mutual information (\textbf{QMI}) so that strongly correlated qubits are placed closer together in the MPS, improving the representation efficiency \cite{araujo_low-rank_2024,jeon_optimal_2024,moritz_convergence_2005,rissler_measuring_2006,barcza_quantum-information_2011,ali_ordering_2021}, and Step 1 is repeated.

In Step 3, we use the MPS for a heuristic preparation of an SMPD (top) \cite{ran_encoding_2020,bohun_scalable_2025} or BMPD (bottom) \cite{mansuroglu_preparation_2025} circuit. Sequential circuits come in multiple variants, marked as left/right/mixed, which refer to the canonical form of the MPS they originate from, and which affect the circuit depth and preparation fidelity. As marked in the "Enhancements" box, each individual nearest-neighbor gate in SMPD circuits can be realized with only 2 CNOTs \cite{shende_minimal_2004,iten_quantum_2016,bohun_scalable_2025} (instead of 3 CNOTs required for a general 2-qubit unitary \cite{vatan_optimal_2004}), and some gates can be reduced to single-body unitaries without affecting the fidelity but saving on the overall gate count \cite{bohun_scalable_2025}. The latter applies also to the BMPD circuits.

Step 4 involves variational optimization of fidelity with the target MPS using the Evenbly-Vidal \cite{evenbly_algorithms_2009,shirakawa_automatic_2024} (top) and Riemannian \cite{luchnikov_qgopt_2021,melnikov_quantum_2023} (bottom) approaches. Optimization is initialized from heuristic circuits from Step 3 and after multiple iterations yields the final solution. In more advanced scenarios, Steps 3 and 4 are performed interchangeably in a loop - heuristic method produces a single new layer of the circuit, the whole circuit is optimized, the heuristic yields a new layer, and so on.

\subsection{Contributions}
This work introduces an end-to-end MPS preparation pipeline, Fig.~\ref{fig:full_pipeline_scheme}, that combines heuristic methods of different origins and circuit structures, with variational optimization techniques, previously considered mostly in isolation. This provides access to highly efficient and scalable state preparation protocols, with the optimal choice of the methods depending on the properties of the target state, required fidelities, gate counts and circuit depths. The pipeline is evaulated on application-relevant classical datasets of varying complexity: Gaussian and L\'evy probability distributions, the chaotic Lorenz attractor \cite{lorenz_deterministic_1963}, and S\&P 500 stock market prices \cite{nugent_sandp500_2018}.

The pipeline incorporates an MPS qubit reordering procedure by placing strongly entangled pairs of qubits at short distances. We formulate this task as a combinatorial quadratic assignment problem (\textbf{QAP}) \cite{koopmans_assignment_1957}.

Heuristic sequential circuits from Ref.~\cite{ran_encoding_2020} are believed to require an exponentially large bond dimension $\tilde{\chi}\sim \mathcal{O}(2^L)$ for a construction of $L$ layers \cite{ran_encoding_2020} (Fig.~4). We argue that the corresponding heuristic circuit can be reliably obtained with only ${\tilde{\chi}\sim \mathcal{O}(\chi) \ll 2^L}$, where $\chi$ is the target MPS bond dimension. This sheds new light on the favorable scaling of the method not only for large systems, but also for many layers.

Lastly, we implement various existing "enhancements" \cite{bohun_scalable_2025} for optimizing the depth and gate counts, see Sec.~\ref{subsec:smpd} for details, and investigate their interplay with the optimization methods.

\paragraph*{}
This article is organized as follows. Sec.~\ref{sec:methods} discusses the pipelines from Fig.~\ref{fig:full_pipeline_scheme} and its individual components. Sec.~\ref{sec:results} shows the pipeline evaluations for various target states and the identification of the most efficient schemes. We conclude in Sec.~\ref{sec:conclusions}.

\section{Methods} 
\label{sec:methods}
The overview of the QSP pipeline from Fig.~\ref{fig:full_pipeline_scheme} is given in Sec.~\ref{subsec:qsppipelineoverview}. Here we motivate and describe each of its constituents in more details, starting with the notation.

\subsection{Notation}
\label{subsec:problem_statement}
Our task is to prepare a quantum state of $N$ qubits, defined by the fixed probability amplitudes ${\psi_{s_1 s_2 \dots s_N} \in \mathbb{C}}$,
\begin{equation}
	\ket{\psi} = \sum_{\lbrace s \rbrace} \psi_{s_1 s_2 \dots s_N} \ket{s_1 s_2 \dots s_N}, \quad \bra{\psi}\ket{\psi} = 1.
	\label{eq:psi}
\end{equation}
by finding a sequence of unitary gates $U_i$ acting on the initial product state $\ket{0}^{\otimes N}$,
\begin{equation}
	\ket{\psi} = \left(\prod_i U_i \right) \ket{0}^{\otimes N},
\end{equation}
forming a corresponding quantum circuit. The construction of unitaries $U_i$ is performed on a classical computer.

\subsection{Matrix Product States}
The central object of our pipeline are quantum states in the MPS form,
\begin{equation}
	\ket{\psi} = \sum_{\lbrace s \rbrace} A^{(s_1)}_1 A^{(s_2)}_2 \dots A^{(s_N)}_N \ket{s_1 s_2 \dots s_N},
\end{equation}
where $A^{(s_i)}_i$ are tensors of dimensions $(2, \chi_i, \chi_{i+1})$, and $\chi=\max_i \chi_i$.
This class of states is chosen for two main reasons. Firstly, the one-dimensional tensor structure, along with the gauge freedom \cite{schollwoeck_densitymatrix_2011}, can be utilized to construct the heuristic approximate SMPD and BMPD circuits with simple two-body nearest-neighbor gates. Secondly, with a weak entanglement one can reach system sizes well beyond those of full state vector simulations. In a typical scenario, quantum hardware takes a classically tractable MPS as an input and increases the state complexity along the circuit execution, potentially exceeding the capabilities of classical computers, giving room for a quantum advantage \cite{jaderberg_variational_2025}. 
Let us note that MPSs can be replaced by more expressive TTNs if one allows for long-range gates and more complex circuits \cite{sugawara_embedding_2025,manabe_state_2025, ballarin_efficient_2025}.

\subsection{Step 1. Compression into MPS}
Construction of the MPS from the probability amplitudes is done in two ways. The standard way is to read \textit{all} $2^N$ probability amplitudes into classical memory and perform successive truncated SVD site by site starting from left(right), yielding an MPS in the right(left) canonical form in $\mathcal{O}(N\chi^3)$ computational complexity \cite{oseledets_tensor_2011,schollwoeck_densitymatrix_2011}. Alternatively, one can construct the MPS using the TCI algorithm \cite{oseledets_tt-cross_2010}, which exploits the low-rank structure and calls for only $\mathcal{O}(N\chi^2)$ adaptively chosen probability amplitudes. With TCI one can efficiently compress very large but low-rank classical datasets, such as smooth functions on finely discretized grids \cite{ritter_quantics_2024}. Appendix~\ref{subsec:TCI} contains a high-level introduction to TCI. Normalization of the MPS, if necessary, requires $\mathcal{O}(N\chi^3)$ operations.

\subsection{Step 2. Reordering qubits}
\label{subsec:reordering}

\begin{figure*}
	\centering
	\includegraphics{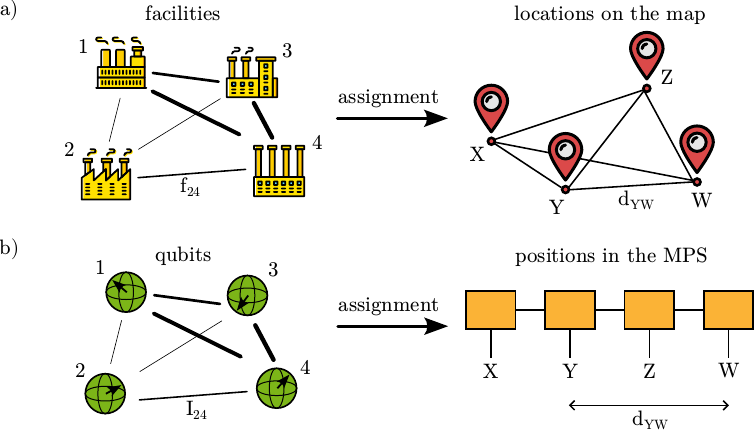}
	\caption{Analogy between a) the standard formulation of the quadratic assignment problem and b) qubit ordering in an MPS. Factories intensively exchanging materials between each other (large flow, marked by bold lines) such as 2 and 4 should be assigned to close cities on the right, just as correlated qubits (large quantum mutual information, marked by bold lines) should be placed close together in the MPS. Numbers enumerate facilities/qubits, capital letters enumerate locations/positions. }
	\label{fig:qap_mps}
\end{figure*}

When two qubits are strongly entangled but placed far apart in the MPS, their entanglement must be “passed” across all intermediate bonds. This affects the required maximal bond dimensions and bond truncation errors, as well as the entanglement at each bond, indicating that placing such a pair of qubits close to each other may yield a more efficient representation of the state, possibly leading to a more efficient QSP. In the context of QSP from MPS, this has been exploited in Refs.~\cite{araujo_low-rank_2024,jeon_optimal_2024}, but both algorithms require exponentially many reconstructions of MPS, making the application to system sizes $N \gtrapprox 10-20$ numerically intractable. 

Interestingly, the problem of finding the optimal ordering of MPS (and other Ansätze) degrees of freedom was addressed earlier in the field of quantum chemistry \cite{moritz_convergence_2005}. Successful approaches are those of Refs.~\cite{rissler_measuring_2006,barcza_quantum-information_2011}, which propose to compute the quantum mutual information between all pairs of single degrees of freedom $i, j$,
\begin{equation}
    I_{ij} = \SvN(\lbrace i \rbrace) + \SvN(\lbrace j \rbrace) - \SvN(\lbrace i \rbrace \cup \lbrace j \rbrace),
\label{eq:qmi}
\end{equation}
where $\SvN (\mathcal{A})=-\Tr \left( \rho_\mathcal{A} \ln \rho_\mathcal{A} \right)$ is the von Neumann entanglement entropy with $\rho_\mathcal{A} = \Tr_{\mathcal{A}'} \ket{\psi}\bra{\psi}$ being the reduced density matrix of subsystem $\mathcal{A}$, and $\mathcal{A}'$ denoting part of the system complementary to $\mathcal{A}$. Loosely speaking, the QMI measures how much uncertainty about subsystem $\lbrace i \rbrace$ is reduced by knowing subsystem $\lbrace j \rbrace$, and vice versa; it quantifies the classical and quantum correlations between $\lbrace i \rbrace$ and $\lbrace j \rbrace$. The QMI matrix can be efficiently computed for the MPS, see e.g. \cite{larrarte_tensor_2025} for details.
The cost function based on QMI and physical distances between degrees of freedom (\textbf{d.o.f.}), in this case orbitals,
\begin{equation}
    C(\pi) = \sum_{i,j} I_{ij} |\pi(i)-\pi(j)|^\eta
\label{eq:cost}
\end{equation}
with $\eta=1, 2$ in \cite{barcza_quantum-information_2011} and $\eta=-2$ in \cite{rissler_measuring_2006}, is used to find the optimal permutation $\pi \in S_N$ of the $N$ d.o.f., which finally leads to a more accurate groundstate energy in the DMRG  \cite{white_density_1992, white_density_1993} calculation. Notice that minimizing (for $\eta>0$) or maximizing (for $\eta < 0$) the cost function amounts to placing orbitals with a large QMI close together. Throughout this work we always use $\eta=1$. It has been rigorously shown that this heuristic directly relates to minimizing the entanglement entropy of MPS \cite{ali_ordering_2021}. In contrast to Refs.~\cite{araujo_low-rank_2024, jeon_optimal_2024}, this approach avoids repeated construction of MPS for different permutations with an excessive number of SVDs, requiring the calculation of QMI only once for a single MPS with a (possibly suboptimal) initial ordering of the d.o.f.. Afterwards, Ref.~\cite{rissler_measuring_2006} uses simulated annealing, while Ref.~\cite{barcza_quantum-information_2011} resorts to an approximate graph optimization method to find the optimal d.o.f. placement. The MPS is then constructed again but with reordered d.o.f.'s and a better groundstate energy in DMRG is achieved.

Our observation is that the problem of optimizing the cost in Eq.~\eqref{eq:cost} is \textit{exactly} equivalent to the quadratic assignment problem \cite{koopmans_assignment_1957}, a well-known NP-hard problem in combinatorial optimization. The QAP is naturally formulated as a facility location problem: given $N$ facilities and $N$ locations, one aims to assign each facility to a location in such a way that the total cost of operation, which depends both on the flow of products or supplies transported between facilities and the distance between locations, is minimized. Formally, the QAP is expressed as
\begin{equation}
    \min_{\pi \in S_N} \sum_{i,j} f_{ij} \, d_{\pi(i)\pi(j)},
\label{eq:qap}
\end{equation}
where $f_{ij}$ is the flow between facilities $i$ and $j$, not necessarily symmetric, and $d_{\pi(i)\pi(j)}$ is the distance between assigned locations $\pi(i)$ and $\pi(j)$.

In our context, one may view the "facilities" as the qubits or other quantum degrees of freedom, and the "flows" $f_{ij}$ as their correlations (e.g., quantified by the quantum mutual information $I_{ij}$), see Fig.~\ref{fig:qap_mps}. The "locations" correspond to the positions in the MPS chain, with the "distances" $d_{\pi(i)\pi(j)}=|\pi(i)-\pi(j)|^\eta$. This makes a direct mapping between the qubit reordering problem in MPS and the QAP, allowing us to use existing heuristic algorithms and software implementations developed over many decades for the latter, see e.g. the \verb|scipy| library implementing the "Fast Approximate QAP" \cite{vogelstein_fast_2015} and "2-opt" \cite{croes_method_1958} optimizers, which we use throughout this work with default settings (in this order). Since $I_{ij}=I_{ji}$, one can use numerical methods specialized in instances of QAP with a symmetric flow.

\subsection{Step 3. Heuristic initialization}
Heuristic methods of mapping the MPS to a quantum circuit are employed to ensure that at least a crude approximation will be found. This is not guaranteed for e.g. a parametrized variational circuit starting from random parameters, whose initial fidelity with the MPS is on average $\mathcal{O}(2^{-N})$. Secondly, heuristic circuits allow for an MPS circuit simulation with an intermediate bond dimension $\tilde{\chi}$ comparable to the MPS bond dimension $\chi$. Finally, if the achieved fidelity is too low, one can use the heuristic result as a warm start initialization for variational optimization. Starting close to a solution mitigates various problems with optimization \cite{rudolph_decomposition_2023}. Below we discuss two heuristic state preparation methods.

\subsubsection{Sequential Matrix Product Disentangler (SMPD)}
\label{subsec:smpd}

\begin{figure*}[t!]
    \centering
    \includegraphics[width=\linewidth]{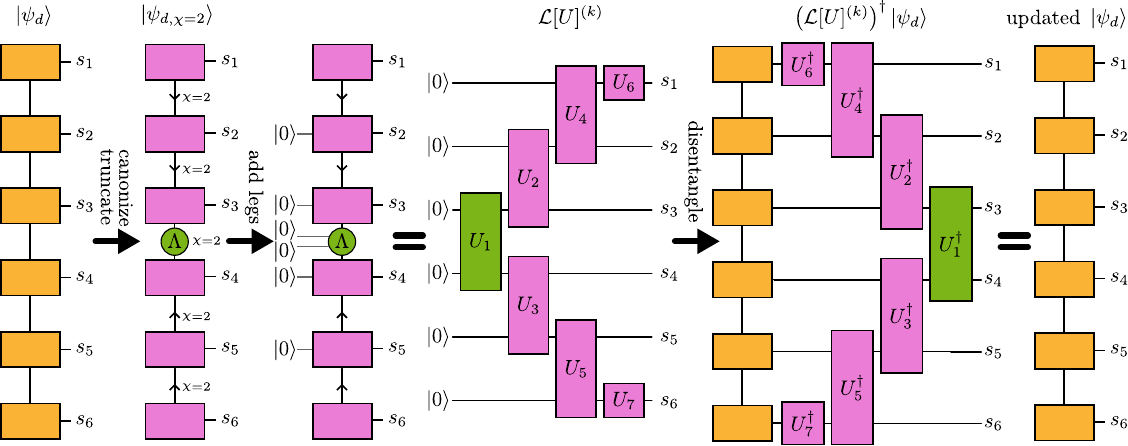}
    \caption{Sequential Matrix Product Disentangler construction, example of the mixed gauge \cite{bohun_scalable_2025}. Initial state $\ket{\psi_d}$ is truncated to $\ket{\psi_{d,\chi=2}}$ and onsite tensors are isometries in the direction depending on the position and gauge (left/right/mixed), with the orthonormality direction denoted by arrows. Fourth "dummy" leg of dimension $1$, corresponding to $\ket{0}$, is added to each tensor with already three legs, giving onsite tensors of dimensions $(1,2,2,2)$, which are isometries when expressed as $2\times4$ matrices. One completes these matrices to $4\times4$ (2-qubit) unitaries $U_i$ by adding two more orthogonal columns with a Gram-Schmidt or QR decomposition-based orthogonalization algorithm. Completing to unitary is not needed for $U_{6,7}$ at both ends which are already $2\times2$ unitaries. Then, the quantum circuit on the right, created by arranging the gates into a layer $\mathcal{L}[U]^{(k)}$ such that the input-output relations of tensors in the MPS are matching with those in the circuit, exactly realizes $\ket{\psi_{d,\chi=2}}$. For the next layer, one applies the inverse of the currently obtained layer to $\ket{\psi_d}$, yielding a new "disentangled" state.
    The operator $U_1$ comes from a diagonal tensor $\Lambda$ and can be realized with $1$ CNOT and one $R_Y$ rotation. Other two-body gates are isometries due to $\ket{0}$ input on one qubit and can be decomposed into $2$ CNOT gates and single qubit rotations. Analogous constructions for left- and right-canonical forms of MPS, which do not contain the $\Lambda$ tensor, can be obtained \cite{ran_encoding_2020}.}
    \label{fig:smpd}
\end{figure*}
This method of an approximate translation of MPS into quantum circuits was proposed by Ran in 2020 \cite{ran_encoding_2020} and widely adopted afterwards \cite{bendov_approximate_2024, rudolph_decomposition_2023, iaconis_tensor_2023, jumade_data_2023, green_quantum_2025,bohun_scalable_2025}. The basic algorithm, a variant of which \cite{bohun_scalable_2025} is presented in Fig.~\ref{fig:smpd}, is based on two observations: i) an MPS with a bond dimension $\chi=2$, when expressed in a canonical gauge with isometric on-site tensors, defines a sequence of two-body nearest-neighbor quantum gates that realize it exactly, and ii) truncation of the bond dimension to $\chi=2$ gives the best rank-2 approximation of the MPS in terms of the Frobenius norm. 

To obtain the first layer of gates in the circuit, one uses the rank-2 approximation of the full MPS $\ket{\psi}$. Then, one can proceed iteratively and construct a next rank-2 representation of a new MPS resulting from the application of the inverse of the previous layer to the original $\ket{\psi}$, "disentangling" it, yielding the next layer of gates. After a maximal number of layers $L$ or the desired fidelity with the product state $\ket{0}^{\otimes N}$ is reached, the process terminates, see Alg.~\ref{alg:smpd}. Layers applied in a reverse order, acting on $\ket{0}^{\otimes N}$, give an approximation of $\ket{\psi}$.

\begin{algorithm}[t!]
\caption{Sequential Matrix Product Disentangler (SMPD)}
\SetKwInOut{Input}{Input}\SetKwInOut{Output}{Output}
\Input{MPS $\ket{\psi}$, max. number of layers $L$, max. intermediate bond dimension $\tilde{\chi}$}
\Output{Layers of QSP circuit $\prod_{k=L}^1 \mathcal{L} \left[ U\right]^{(k)}$}

\BlankLine
$k \gets 1$\;
$\ket{\psi_d} \gets \ket{\psi}$ - "disentangled state"\;
\While{$k\leq L$}{
  Compute the norm $\mathcal{N} =\left|\bra{\psi_d}\ket{\psi_d}\right|^2$\;
  Truncate $\mathcal{N}^{-1/2} \ket{\psi_d}$ to $\ket{\psi_{d,\chi=2}}$ in left-, right- or mixed-canonical form with SVDs, renormalizing singular values to $\lambda_1^2+\lambda_2^2=1$ at each bond\;
  Transform $\ket{\psi_{d,\chi=2}}$ into a layer of gates $\mathcal{L}\left[U\right]^{(k)}$ as in Fig.~\ref{fig:smpd}\;
  Disentangle current state for the next step, keeping a maximal bond dimension $\tilde{\chi}$,
  $\ket{\psi_d} \gets \left( \mathcal{L}\left[U\right]^{(k)}\right)^\dagger\ket{\psi_d}$\;
  $k \gets k+1$\;
}
\label{alg:smpd}
\end{algorithm}

The main advantage of this procedure is the deterministic formulation. At each layer, the unitary gates are uniquely determined by the truncation of MPS to $\chi=2$, leading to a refined approximation. Only the nearest-neighbor gates are required, minimizing the need for SWAP operations on physical platforms with nearest-neighbor connectivity.

Nevertheless, one can identify several potentially weak points. The most important one is the circuit depth scaling as $\mathcal{O}(N)$ with the number of qubits $N$. It is better than the naive $\mathcal{O}(2^N)$ scaling for a general QSP, but may still pose a challenge in a practical realization with $10$s of qubits. Another drawback is the empirical observation that after a few layers are applied, $L\sim 5-10$, the fidelity of the circuit typically starts to saturate. 

Another potential drawback brought up in Ref.~\cite{ran_encoding_2020} is the need for an exponentially large bond dimension $\tilde{\chi} \sim \mathcal{O}(2^L)$ for the construction of $L$-layer SMPDs, limiting classical calculations to a few layers only. In Sec.~\ref{subsec:smpd_convergence} we independently revisit the calculations of Ref.~\cite{ran_encoding_2020} and find that deep SMPD circuits are simulable with bond dimensions $\tilde{\chi} \sim \mathcal{O}(\chi) \ll 2^L$.

\begin{figure}[t!]
    \centering
    \includegraphics[width=\columnwidth]{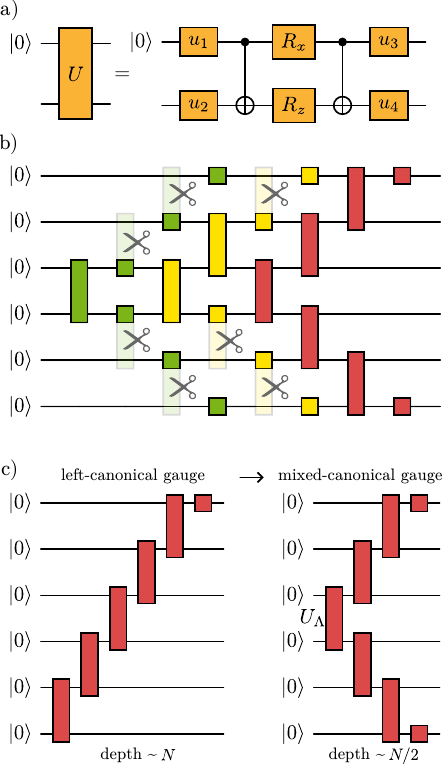}
    \caption{Enhancements of the SMPD. 
    	a) Isometry is decomposed into 2 CNOTs and single-qubit gates.
    	b) If the MPS bond is already disentangled, apply only single-qubit gates.
    	c) Using the mixed-canonical gauge decreases depth in half.}
    \label{fig:enhancements_smpd}
\end{figure}

Building on the results of Ref.~\cite{bohun_scalable_2025}, we employ three "enhancements" of the SMPD implementation that lead to either lower gate counts or shallower circuits, as illustrated in Fig.~\ref{fig:enhancements_smpd}.

The first enhancement, Fig.~\ref{fig:enhancements_smpd}a), is based on the observation that every SMPD layer consists of two-body gates acting on one qubit in a fixed state $\ket{0}$, and the other qubit in an arbitrary state. Thus, the gate is defined by only 2 out of all 4 orthogonal columns of the 2-qubit unitary matrix, i.e. it is an isometry from 1 to 2 qubits. It is known that even though an exact decomposition of a 2-qubit unitary requires exactly 3 CNOT gates in the general case \cite{vatan_optimal_2004}, the decomposition of a 1-to-2 qubit isometry can be performed with only 2 CNOTs \cite{iten_quantum_2016}. We implement a decomposition of both unitaries and isometries into elementary single-qubit rotations and CNOT gates using an Ansatz from Ref.~\cite{shende_minimal_2004}. We stress that the single-qubit rotation angles are deterministically calculated from the matrix elements of the decomposed operator, not variationally optimized, see Appendix~\ref{app:isometry} for details. 

The second enhancement, shown in Fig.~\ref{fig:enhancements_smpd}b), exploits the spatial structure of entanglement in the MPS. If a certain bond has already been disentangled after applying the disentangling operators, i.e. it reached a bond dimension of $1$ after truncation at a desired singular value cutoff level $\varepsilon_{\text{SVD}}$, there is no need to apply a two-body gate around this bond. One can apply only a single qubit rotation, significantly reducing the overall number of two-body gates needed to construct the circuit.

Finally, it is also beneficial to consider the SMPD for an MPS expressed in the mixed gauge, as in Figs.~\ref{fig:smpd}. That is, one chooses a certain bond as a center of orthogonality, transforms all tensors to the left into right-orthogonal, and all tensors to the right into left-orthogonal form, and truncates the bond dimension to $\chi=2$. This results in a circuit layer with gates in the "V"-shape, with a depth smaller by up to 50\% than from the left- or right-canonical gauge, cf. Fig.~\ref{fig:enhancements_smpd}c). The central bond with two singular values $\Lambda_1 \geq \Lambda_2 \geq 0$, normalized to $\Lambda_1^2 + \Lambda_2^2 =1$, corresponds to an isometry from $\ket{00}$ state to a 2-qubit state. Notice that the bond tensor with singular values is a diagonal gate $U_1\equiv U_\Lambda$ defined by $U_\Lambda \ket{00} = \Lambda_1 \ket{00} + \Lambda _2\ket{11}$. Such a gate can be realized with only a single $R_Y(\theta)$ rotation and one CNOT gate. Specifically, assuming
\begin{equation}
R_Y(\theta) = e^{-i \frac{\theta}{2} Y} 
= \begin{bmatrix}
\cos \tfrac{\theta}{2} & -\sin \tfrac{\theta}{2} \\
\sin \tfrac{\theta}{2} & \cos \tfrac{\theta}{2}
\end{bmatrix}
\end{equation}
with Pauli gate $Y\equiv \sigma^y$, one can choose ${\theta^* = 2 \arctan \left(\Lambda_2 / \Lambda_1 \right)}$ such that 
\begin{equation}
    U_\Lambda = \text{CNOT} \cdot (R_Y(\theta^*) \otimes \mathbb{1}).
\end{equation}
This construction gives explicit parameters for an exact realization of the central gate, requires only a single CNOT, and avoids variational optimizations. This is in contrast to Ref.~\cite{bohun_scalable_2025}, where the first disentangler layer also has $U_\Lambda$ realized with a single CNOT, but for further layers it is optimized variationally from a 2-CNOT Ansatz with single qubit rotations to minimize the R\'enyi entropy at the central bond. Both other enhancements described in this section also apply to the mixed-gauge SMPD.

For completeness, let's note that an analogue of the SMPD for TTN was introduced in Ref.~\cite{sugawara_embedding_2025}, and an open source Python implementation of the basic version of the SMPD algorithm for \verb|qiskit| is available \cite{millar_mps-to-circuit_2025}.

\subsubsection{Brickwall Matrix Product Disentangler (BMPD)}
\label{subsec:bmpd}

\begin{figure*}
    \centering
    \includegraphics[width=\linewidth]{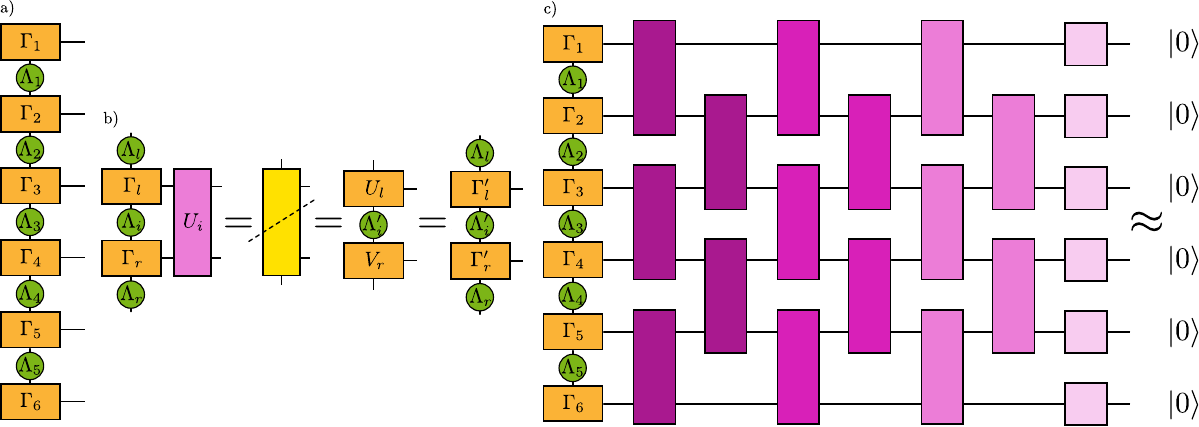}
    \caption{a) The MPS is transformed into the $\Gamma$-$\Lambda$ gauge, also known as the Vidal's gauge. Singular values stored at each bond allow for a direct entropy calculation. b) Application of a gate $U_i$ which is supposed to decrease R\'enyi entanglement entropy at the $i$-th bond. Contraction of the network is followed by SVD of the resulting tensor across a dashed line. New $\Gamma_{l,r}'$ matrices are obtained as $\Gamma_l'=\Lambda_l^{-1}U_l$ and $\Gamma_r'=V_r\Lambda_r^{-1}$. The MPS stays in the $\Gamma$-$\Lambda$ canonical form and only the $\Gamma_l$, $\Gamma_r$, $\Lambda_i$ tensors are affected by the gate application. c) Layer-by-layer disentangling of an MPS. Parameters of gates in a single column are variationally optimized in parallel to minimize R\'enyi entropy at each bond. After optimization, the gates are contracted with the MPS, resulting in a lower bond dimension on average, and the optimization proceeds to the next column of gates. The last layer consists of single-body rotations. Hermitian conjugate of this quantum circuit prepares an approximation of the MPS from the $\ket{0}^{\otimes N}$ initial state.}
    \label{fig:bmpd}
\end{figure*}

An alternative heuristic state preparation algorithm has been proposed by Mansuroglu and Schuch \cite{mansuroglu_preparation_2025}. The construction is based on the MPS representation in the $\Gamma$-$\Lambda$ gauge, first introduced by Vidal \cite{vidal_efficient_2003}, in which the singular values of the Schmidt decomposition are stored as diagonal $\Lambda$ matrices at each bond, along with $\Gamma$ tensors at each site,
\begin{equation}
    \ket{\psi} = \sum_{\lbrace s \rbrace} \Gamma_1 \Lambda_1 \Gamma_2 \Lambda_2 \dots \Gamma_{N-1}\Lambda_{N-1}\Gamma_N\ket{s_1 s_2 \dots s_N},
\label{eq:gammalambda}
\end{equation}
see Fig.~\ref{fig:bmpd}a) for a graphical representation. 
This form can be obtained by first bringing the MPS into left- or right-canonical form, and then performing a sequence of SVDs \cite{schollwoeck_densitymatrix_2011}.\footnote{Bringing the MPS into left- or right-canonical form first is necessary when we start with an MPS not expressed in any of these two gauges. If we start from a dense state vector, the transformation into MPS in the $\Gamma$-$\Lambda$ gauge can be performed directly, see \cite{schollwoeck_densitymatrix_2011}.}

The $\Gamma$-$\Lambda$ gauge is chosen due the possibility to simultaneously access Schmidt values at each bond, which is not the case e.g. for the mixed gauge where Schmidt values only at a single bond can be read directly, and one has to move the center of orthogonality in order to access Schmidt values at different bonds. 

The core routine of the algorithm is the application of a two-body unitary $U_{i,i+1}$ on sites $i, (i+1)$, which is supposed to decrease the entanglement entropy at the $i$-th bond. Parameters of the unitary are variationally optimized so that the $\alpha$-order R\'{e}nyi entropy,
\begin{equation}
    \mathcal{S}_{R,\alpha, i} = \frac{1}{1 - \alpha} \log \left( \sum_k (\Lambda_i)_k^{2\alpha}\right)
\label{eq:renyi}
\end{equation}
where $(\Lambda_{i})_k$ is the $k$-th singular value at bond $i$, is minimized. An Ansatz for a general disentangling unitary has the following form:
\begin{eqnarray}
    U_{i,j} &(\theta) &= e^{-i\theta_{i,1} X_i X_j}e^{-i\theta_{i,2} Y_i Y_j}e^{-i\theta_{i,3} Z_i Z_j} \\\nonumber
    &\times& \left( e^{-i(\theta_{i,4} X_j + \theta_{i,5} Y_j + \theta_{i,6} Z_j)} \right.\\\nonumber
    &&\quad \otimes 
    \left. e^{-i(\theta_{i,7} X_i + \theta_{i,8} Y_i + \theta_{i,9} Z_i)}\right)
\label{eq:general_disentangler}
\end{eqnarray}
which is a decomposition of a general two-qubit SU(4) matrix \cite{kraus_optimal_2001} without a pair of single-qubit gates at the end which do not influence the entanglement.

Application of the disentangling unitary to an MPS is illustrated in Fig.~\ref{fig:bmpd}b).\footnote{In \cite{mansuroglu_preparation_2025} the unitary $U_i$ is first decomposed with SVD into two tensors for contraction efficiency, which we omit. Our implementation uses autodifferentiation instead of the parameter-shift rule for obtaining gradients, which requires $1$ SVD and its backpropagation instead of $18$ SVDs.} Each gate application preserves the $\Gamma$-$\Lambda$ gauge, and modifies only local tensors $\Gamma_l$, $\Gamma_r$, $\Lambda_i$. This allows for an efficient parallelization of the disentangling process over gates applied to pairs of neighboring qubits around every second bond. After disentangling the odd bonds, one proceeds with another set of unitary gates to disentangle the remaining even bonds. This gives two sublayers, which form a layer. Iterating the process gives a disentangling quantum circuit with a brickwall structure, see Fig.~\ref{fig:bmpd}c). After many such disentangling layers are applied, one is left with a state with a small R\'{e}nyi entropy, i.e. approximately a product state. Then, a layer of single-body rotations allows one to reach the state $\ket{0}^{\otimes N}$. Parameters of these unitaries result from truncation of the final MPS to bond dimension $\chi=1$. If the full disentangling circuit corresponds to a unitary $U$, the quantum state can be prepared by inverting it: $\ket{\psi} \approx U^\dagger \ket{000\dots0}$.

There are several major advantages of the BMPD, as underlined in \cite{mansuroglu_preparation_2025}:

i) Instead of optimizing all gates at once, which would be exponentially costly due to an expected increase of the MPS bond dimension after each layer, one performs a tractable layer-by-layer optimization, keeping the MPS form. After each layer, one is left with a less entangled state, so the MPS bond dimension decreases on average. This makes a classical simulation of deep circuits possible. 

ii) Since the variational optimization for each gate is strictly local
the loss landscape provably doesn't suffer from the usual problem of barren plateaus, making the method scalable to large systems. 

iii) The truncation error made during disentangling of the $i$-th bond affects only the tensors $\Gamma_i$, $\Lambda_i$, $\Gamma_{i+1}$. Theoretical guarantees for the total state preparation error assuming a fixed maximal bond dimension are provided in \cite{mansuroglu_preparation_2025}. 

iv) In principle, the circuit depth is proportional to the number of disentangling layers and independent of the system size. However, our observation is that if we want to disentangle qubits separated by a distance $\xi$, since a single disentangler acts only on a pair of neighboring sites, we need a circuit depth of at least $\xi/2$, i.e. linear in the system size and the same as SMPD in the mixed gauge if $\xi\sim N$. This is a consequence of the Lieb-Robinson bounds on the propagation of quantum correlations \cite{lieb_finite_1972}, see also \cite{bravo_prieto_scaling_2020}(Sec.~3).

v) The circuit Ansatz involves only nearest-neighbor gates, which is an advantage for many NISQ platforms such as those with trapped ions \cite{cirac_quantum_1995} or superconducting qubits \cite{nakamura_coherent_1999}.

vi) The Ansatz for a parametrized disentangling gate can be either a general disentangler, as in Eq.~\eqref{eq:general_disentangler}, or a hardware-efficient gate, depending on the native hardware capabilities.

The hyperparameter ${\alpha>0}, {\alpha \neq 1}$, the exponent in the R\'enyi entropy formula, Eq.~\eqref{eq:renyi}, needs to be appropriately chosen. Small $\alpha \rightarrow 0$ favors minimization of the Schmidt values across many orders of magnitude, leading to an exploration of a large space of entangled states to find a superposition leading to small total Renyi entropy. On the contrary, for a large $\alpha \rightarrow \infty$, the entropy reads $S_{\infty,i} = - \log ( (\Lambda_i)_1^2)$, and is dependent only on the leading singular value. In this case, the minimization leads to a domination of the leading singular value, asymptotically reaching $1$, and suppression of the other singular values. This decreases the maximal bond dimension, making the classical simulation more efficient. Thus, one needs to find a sweet spot between the exploration and numerical cost. Throughout this work we choose $\alpha=2$ as an intermediate case.

Finally, having in mind the limitations of NISQ devices such as limited achievable gate counts, one can cease adding new disentangling gates if a desired small level of R\'enyi entropy and/or singular value magnitude at a given bond is reached. This is in full analogy to the enhancement applied to SMPD and depicted in Fig.~\ref{fig:enhancements_smpd}b). In this way, one avoids a physical implementation of gates possibly very close to an identity (up to single-qubit rotations) whose application unnecessarily takes up quantum resources and introduces errors.

\subsection{Gate optimization algorithms}
\label{subsec:gate_optimization}

Let us consider a quantum circuit which realizes a state $U\ket{0}^{\otimes N}$ through a sequence of $M$ few-body unitary gates,
\begin{equation}
	U = U_M U_{M-1}...U_2 U_1.
\end{equation}
Our goal is to find gates $U_i$ that maximize the absolute value of the fidelity with a target state $\ket{\psi}$,
\begin{equation}
	F = \bra{000\dots 0}U^\dagger \ket{\psi},
	\label{eq:F}
\end{equation}
under constraint that the gates are unitary. 

\subsubsection{Evenbly-Vidal sweeping optimization algorithm}
\begin{figure}[t!]
    \centering
    \includegraphics[width=\columnwidth]{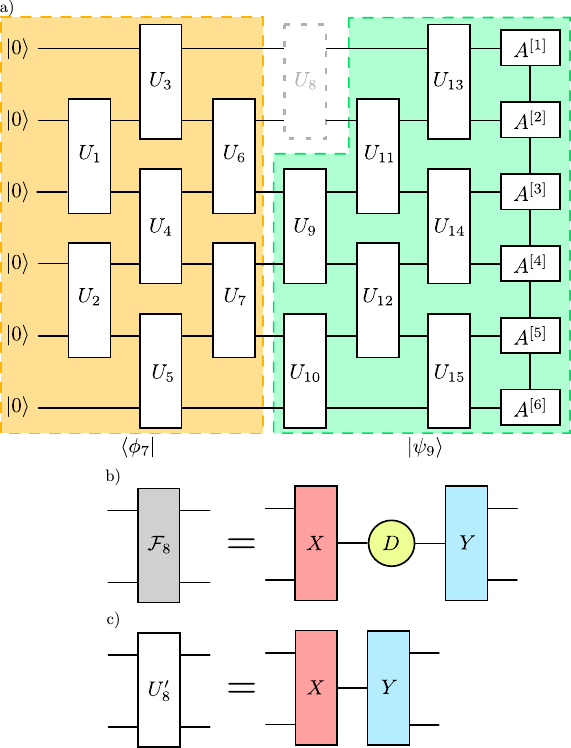}
    \caption{Step of the Evenbly-Vidal optimization, maximizing fidelity of the circuit with the MPS. a) Environment $\mathcal{F}_m$ calculation for $m=8$. The entire tensor network is contracted except for the $U_8$ gate. b) SVD of the environment tensor. c) Optimally updated gate $U_8$.} 
    \label{fig:evenblyvidal}
\end{figure}
One can simplify the problem and concentrate on finding a single optimal gate $U_m$ while keeping the other gates fixed, with a method introduced in 2009 by Evenbly and Vidal \cite{evenbly_algorithms_2009} for isometric tensor networks, and later readapted by Shirakawa et al. for quantum circuits \cite{shirakawa_automatic_2024}. In order to compute an update for a single gate, one computes its \textit{environment tensor} $\mathcal{F}_m$ by contracting the tensor network representing fidelity with the target state with the gate of interest removed, see Fig.~\ref{fig:evenblyvidal}. The new, optimally updated gate $U_m'$ is then given by the SVD of $\mathcal{F}_m$ with singular values replaced by $1$, see Appendix~\ref{app:ev} for details. Visiting all $U_m$ twice, going from the first gate to the last and back, makes a single \textit{sweep}, which can be repeated $\nswpEV$ times until convergence. During the sweep, the state to the left and right of the current gate can be kept in the MPS form.

As noticed in Ref.~\cite{rudolph_decomposition_2023}, the EV optimizer gives a strong update, which can decrease the efficiency of the full circuit optimization. Therefore, it was proposed to apply a finite learning rate $0<\beta\leq1$, so that the update  $U_m''$ reads
\begin{equation}
    U_m'' = U_m (U_m^\dagger U_m')^\beta.
    \label{eq:learning_rate}
\end{equation}
In case $\beta=1$, $U_m''=U_m'$; with decreasing $\beta>0$ the update becomes smaller. The matrix power can be computed numerically using the eigendecomposition of the matrix and a standard exponentiation of the diagonal eigenvalue matrix. In Appendix~\ref{app:learningrate} we study the final fidelity of the optimized states for two different datasets and find a rather limited variation of the result in the range $\beta\in[0.2, 1]$. Throughout this paper we choose an intermediate $\beta=0.6$, same as in Ref.~\cite{rudolph_decomposition_2023}.

\subsubsection{Riemannian variational optimization}
A standard approach to variational quantum circuit optimization with classical computers is to parametrize the gates in terms of e.g. rotation angles $\exp(-i \theta_{j,\alpha} \sigma_\alpha)$ and update parameters $\theta$ until convergence of the cost function such as energy or fidelity. However, it is known that variational quantum circuits suffer from a problem of barren plateaus with parametrizations of this kind, i.e. the cost function is (approximately) constant almost everywhere in parameter space, except for exponentially (in the system size $N$) narrow regions where it reaches a minimum. 

One of the ways to partially mitigate this effect is to avoid parametrizations in terms of angles in the exponential, and directly optimize the matrix elements of the $K$-body $2^K \times 2^K$ unitary gates (in our case, $K\in\lbrace1,2\rbrace$), which imposes the unitarity constraint. In such a case of a constrained and curved variational manifold, one can apply Riemannian optimization techniques \cite{luchnikov_qgopt_2021,melnikov_quantum_2023}. To stay on the manifold during gradient-based optimization and to perform updates in accordance with its local geometry, one projects the Euclidean gradient $\nabla f(u)$ onto the tangent space to get the Riemannian gradient $\nabla_Rf (u)$, and generalizes the Euclidean point and vector transport. Details are provided in Appendix~\ref{app:riemannian}.

Here we use the Riemannian version of the Adaptive Moment Estimation (Adam) first-order gradient optimizer \cite{kingma_adam_2017} implemented in the  \verb|QGOpt| software package \cite{luchnikov_qgopt_2021}. The loss function is the fidelity of the circuit with the target MPS from Eq.~\eqref{eq:F}, and we optimize over complex matrix elements of unitary gates $U_1,\dots U_M$. Contraction of the corresponding tensor network defines a computational graph which allows for autodifferentiation to obtain gradients of variational parameters \cite{liao_differentiable_2019}. For tensor network contractions, which take the most of the computational time and are exponentially costly in general, we utilize the \verb|cotengra| hyperoptimized contractions \cite{gray_hyper_2021}.

\section{Results}
\label{sec:results}
We perform numerical evaluations s of different variants of the QSP pipeline described in Sec.~\ref{sec:methods}. We start in Sec.~\ref{subsec:smpd_convergence} by demonstrating that moderate bond dimensions $\tilde{\chi} \ll 2^L$ in the classical simulation are sufficient to construct $L$-layer SMPD circuits, revisiting the results of Ref.~\cite{ran_encoding_2020} on the same input data. Then, in Sec.~\ref{subsec:datasets} we introduce four classical datasets which are amplitude-encoded as quantum states for QSP benchmarking. Sec.~\ref{subsec:results_reordering} is devoted to our qubit reordering algorithm analysis. Results from evaluating the QSP methods based on disentanglers are presented in Sec.~\ref{subsec:results_disentanglers} (only disentanglers) and Sec.~\ref{subsec:results_optimized} (disentanglers combined with optimization).

\subsection{State preparation with deep layered SMPD}
\label{subsec:smpd_convergence}

A potential problem with the SMPD method brought up in Ref.~\cite{ran_encoding_2020} is the need for an exponentially large bond dimension $\tilde{\chi}\sim \mathcal{O}(2^L)$ for the construction of $L$-layer disentanglers, limiting classical calculations to a few layers only, above which the fidelity rapidly decays. The argument is based on the opposite error propagation directions for obtaining the circuit and for executing it to prepare the state, and analogies to the TEBD time evolution algorithm. Here, using the same input data, we demonstrate no such decays of fidelity occur even for deep circuits.

We consider the 1D transverse field Ising Hamiltonian with open boundary conditions,
\begin{equation}
	H = \sum_{n=1}^{N-1} S_n^z S_{n+1}^z-h_x\sum_n S_n^x,
	\label{eq:1dtfim}
\end{equation}
with $S_n^{x,y,z}=\sigma_n^{x,y,z}/2$ denoting spin-1/2 operators and $\sigma_n^{x,y,z}$ Pauli matrices at site $n$. We use the DMRG algorithm \cite{white_density_1992, white_density_1993} to obtain the groundstate for $N=48$ spins close to the critical point at $h_x=0.5$, with the absolute energy tolerance of $10^{-12}$, SVD singular value cutoff $\epsilon = 10^{-13}$ (relative to the maximal singular value), maximal bond dimension $\chi=64$ and double floating point precision (complex128 data type), implemented in \verb|quimb| \cite{gray_quimb_2018}. The energy converges to $E_0=-15.188704241243$ after $10$ DMRG sweeps, yielding an MPS with bond dimension $\chi=25$. 


In \cite{ran_encoding_2020}(Fig.~4) the same state is considered for state preparation with the SMPD algorithm. Through the negative logarithmic fidelity per site,
\begin{equation}
	F_L = - \frac{\ln |\bra{\psi} \prod_{k=L}^1 \mathcal{L} \left[ U\right]^{(k)} \ket{0}^{\otimes N}|}{N}
\label{eq:NLF}
\end{equation}
calculated for circuits with a varying number of layers $L$ and bond dimensions used for their construction $\tilde{\chi}$, it is shown that $F_L$ abruptly increases whenever $L \gtrapprox \log_2 \tilde{\chi}$, reaching values on the order of $F_L\approx0.5$, which for this system size translates to a "global" infidelity $\Ical = 1-  |\bra{\psi} U^\dagger_L \cdots U^\dagger_2 U^\dagger_1 \ket{0}|^2 \approx 1 - 10^{-21}$, i.e., failure to prepare the state.

 \begin{figure}[t!]
 	\centering
 	\includegraphics[width=\columnwidth]{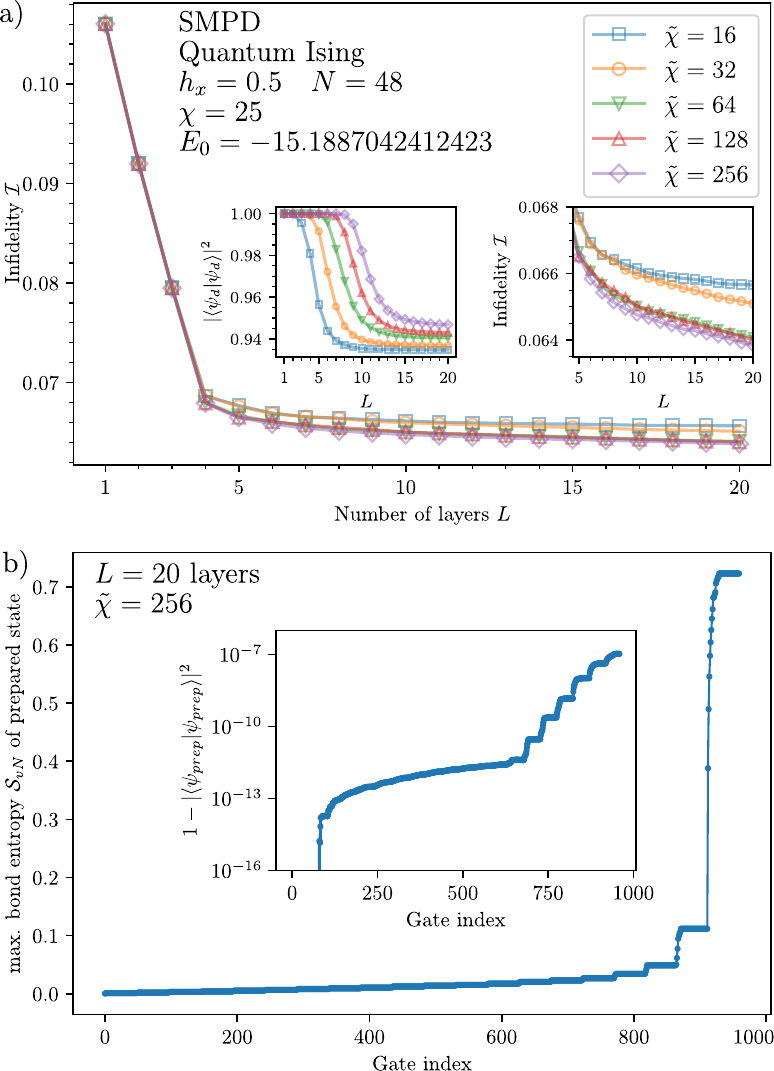}
 	\caption{a) Infidelity of the 1D quantum Ising model groundstate preparation for a given number of SMPD layers. There is no abrupt increase of infidelity at $L\approx \log_2 \tilde{\chi}$. For $L>5$, the improvement with increeasing $L$ is slow but noticeable (right inset). Truncation errors of the disentangled state $\ket{\psi_d}$ result in a decreased state norm (left inset). The infidelities were computed using MPS simulator with bond dimension $\tilde{\chi}'=256$. b) Maximal bond entanglement of the prepared state after each gate application ($49$ gates per layer). Most entanglement is created in the last layer, so at early stages the circuit can be simulated efficiently with a small bond dimension. Inset shows error in the norm of the prepared state which stays exponentially small.}
 	\label{fig:smpd_convergence_twopanels}
 \end{figure}

Our calculation of infidelity $\Ical$ as a function of the number of SMPD layers $L$ for various choices of $\tilde{\chi}$ is shown in Fig.~\ref{fig:smpd_convergence_twopanels}a). A single layer is enough to approximate the state with infidelity $\Ical\approx 0.106$. A significant improvement when adding new layer is visible for the first $4-5$ layers. Although extending the circuit further systematically decreases the infidelity, the improvement per layer slows down, see the right inset. The difference from the result of Ref.~\cite{ran_encoding_2020} is clear: there is no abrupt increase of the infidelity to $\Ical \rightarrow 1$ after the number of layers exceeds $L \gtrsim \log_2 \tilde{\chi}$. In fact, even for $\tilde{\chi}=16$, which does not allow the prepared state to reach the target $\chi=25$, we still observe values of infidelity very close to those for $\tilde{\chi}=256$. To ensure that the presented small infidelities did not result from the truncations in the QSP circuit MPS simulation, even though the disentangler circuits were constructed with $\tilde{\chi}$ each, we used bond dimension $\tilde{\chi}' = 256$ to compute the infidelities. We verified that the change of the squared norm of the \emph{prepared} state 
\begin{equation}
	\ket{\psi_{\text{prep}}}=\prod_{k=L}^1 \mathcal{L} \left[ U\right]^{(k)} \ket{0}^{\otimes N}
\end{equation} did not exceed $1.5 \times 10^{-7}$ in all cases. The plot was constructed for SMPD resulting from right-canonical MPS gauge to match Ref.~\cite{ran_encoding_2020}, but we observed the same behavior for both left- and mixed-canonical form-based SMPDs. Not aiming for an efficient preparation but for a comparison, we did not use any of the SMPD algorithm enhancements from Fig.~\ref{fig:enhancements_smpd}. 

The left inset of Fig.~\ref{fig:smpd_convergence_twopanels}a) shows the squared norm of the \emph{disentangled} state
\begin{equation}
	\ket{\psi_d} = \left(\prod_{k=L}^1 \mathcal{L} \left[ U\right]^{(k)}\right)^\dagger \ket{\psi}
\end{equation}
after $L$ disentangling layers applied to the target MPS $\ket{\psi}$ with a fixed $\tilde{\chi}$. It is clear that the norm drops noticeably from $1.0$ to $0.93-0.95$ for the deepest circuits due to truncation errors, and the drop starts to occur at roughly $\tilde{\chi}\approx 2^L$. Nevertheless, there is still enough "signal" in this truncated state to construct further layers because the state preparation infidelity is systematically improving, as demonstrated in the right inset. The improvement of infidelity per layer is very slow, but we attribute this effect to the specific choice of the target state. 
In one of the examples coming from classical datasets, we observe that infidelity can drop from $\Ical=0.7$ to $\Ical=0.01$ across $L=300$ SMPD layers, see further Sec.~\ref{subsec:results_disentanglers} and Fig.~\ref{fig:results_disentanglers}.

Fig.~\ref{fig:smpd_convergence}b) demonstrates efficient classical simulability of the state preparation SMPD circuits. We compute the maximal entanglement entropy in the system after each gate application in the state preparation process. The system starts from a product state $\ket{0}^{\otimes N}$ with zero entanglement. Then, the entanglement grows with each gate/layer, but the growth is very slow and the system is classically simulable with a small bond dimension, see the inset for the accumulated truncation error. Most of the entanglement is produced in the last few SMPD layers. On the one hand, this is a positive message about the classical simulability of SMPD circuits, making their state preparation circuit results verifiable on classical computers. On the other hand, the circuit spends most of the time applying very weakly entangling gates, signaling that the state preparation scheme is far from optimal.

Concluding, we do not confirm the results of \cite{ran_encoding_2020}(Fig.~4), implying that the required bond dimension for SMPD construction is exponential in the number of layers, ${\tilde{\chi}=2^L}$. We find that both the SMPD construction, and the following state preparation, are classically simulable with $\tilde{\chi} \ll 2^L$, making the method useful also for more demanding scenarios requiring $L\gg 1$ layers. Further study is performed in Appendix~\ref{app:smpd_convergence}, confirming that the rank-2 approximation is robust to truncation errors, there are no noticeable infidelity gains for $\tilde{\chi}\geq 50$, and the conclusions made here also hold for more complex states such as random MPS.

\subsection{Classical datasets}
\label{subsec:datasets}
\begin{figure*}[t!]
    \includegraphics[width=\linewidth]{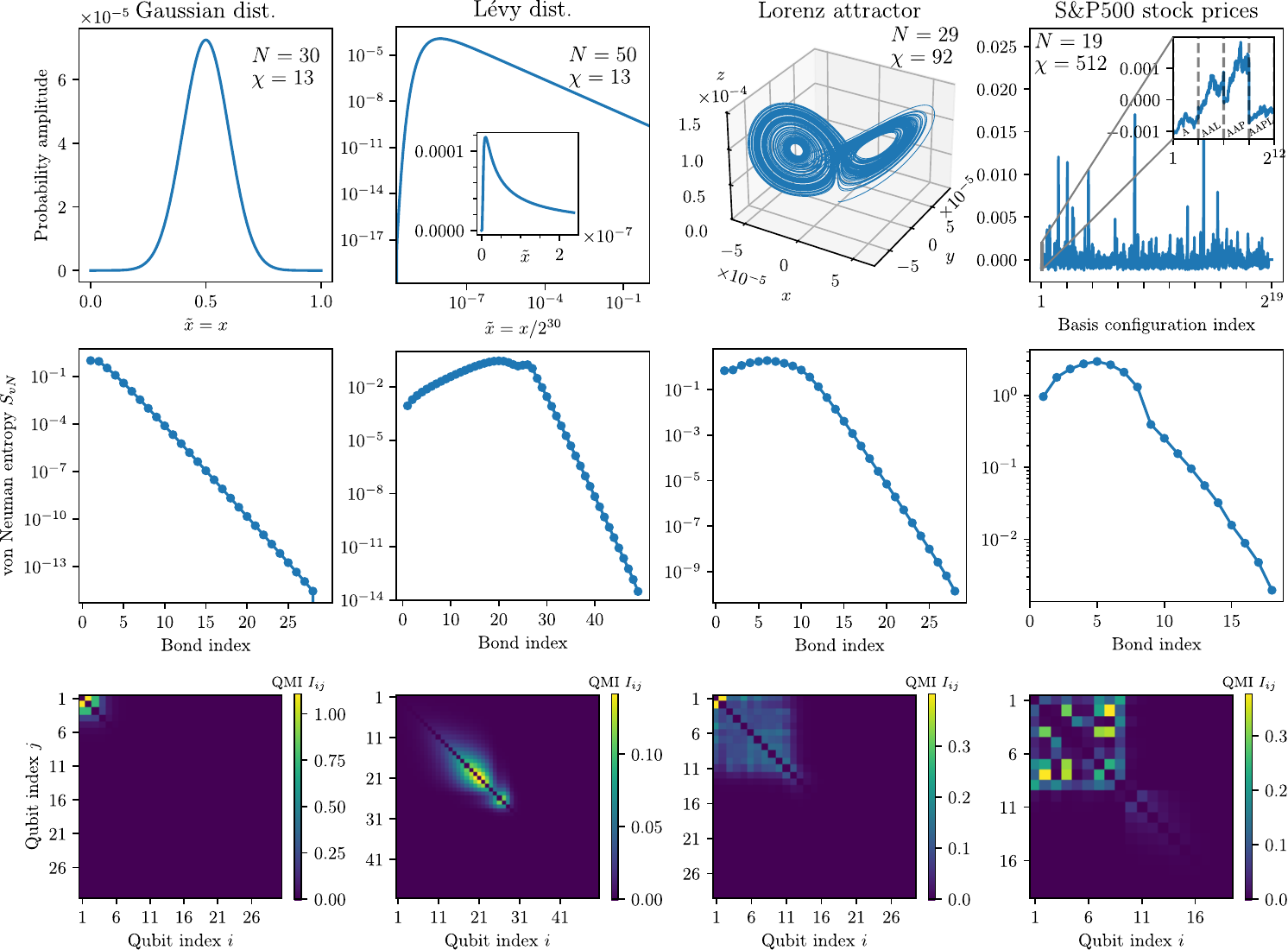}
\caption{Classical datasets used for benchmarking the quantum state preparation methods. Top row: Probability amplitudes encoded with an MPS on $N$ qubits with bond dimension $\chi$, either with TCI (Gaussian, L\'evy distributions) or with direct step-by-step SVD compression of a dense state vetctor (Lorenz attractor, S\&P 500). Insets zoom into parts of the larger plots. In the S\&P 500 case, gray dashed lines are put every $2^{10}=1024$ points and distinguish the companies A, AAL, AAP, AAPL, etc. arranged in alphabetical order. Middle row: Entanglement entropy at MPS bonds. Exponential decay starting from an index $k$ representing the relevant lengthscale of the problem $\xi=1/2^k$ is observed, see text. Bottom row: Quantum mutual information between all pairs of qubits.  }

\label{fig:datasets}
\end{figure*}

For benchmarking the state preparation methods in the context of their future potential practical applications, we select four classical datasets, all presented in Fig.~\ref{fig:datasets}. 

\subsubsection{Probability distributions}
We encode two continuous univariate probability distributions $f(x)$ into quantum probability amplitudes ${\psi(x)=\sqrt{f(x)}}$. As a baseline, we choose the normal distribution
\begin{equation}
    f(x) = \frac{1}{\sqrt{2\pi \sigma^2}} e^{-(x-\mu)^2/(2\sigma^2)}
 \label{eq:gaussian}
\end{equation}
which is a fundamental object in statistics due to the central limit theorem \cite{montgomery_applied_2014}. At the same time, as it later turns out, it is easy to prepare as a quantum circuit, so that we can check the convergence of our algorithms. We set $\mu=0.5$ and $\sigma=0.1$. For an interested reader, we recommend Refs.~\cite{iaconis_quantum_2024, xie_efficient_2025} for an in-depth analysis of univariate normal distribution preparation on quantum computers, as well as \cite{manabe_state_2025, ballarin_efficient_2025} for recent advances on multivariate Gaussian preparation.

The second example, inspired by Ref.~\cite{bohun_scalable_2025}, is the L\'{e}vy distribution
\begin{equation}
    f(x) = \sqrt{\frac{c}{2\pi}} \frac{e^{-c/(2x)}}{x^{3/2}}, \quad x>0
 \label{eq:levy}
\end{equation}
where $c>0$ is a scale parameter. It is a heavy-tailed, scale-free distribution, resulting from additive contributions at multiple different scales. The L\'{e}vy distribution appears in anomalous diffusion processes like L\'{e}vy flights \cite{chechkin_introduction_2008}, where occasional very long jumps distinguish the behavior from standard Brownian motion. It also emerges in critical phenomena near phase transitions, quantum mechanics with long-range interactions, as well as in biological systems \cite{zaburdaev_levy_2015}. Properties of financial data, especially in the context of risk management where rare but extreme events play an essential role, are also often captured by the L\'{e}vy distribution \cite{bouchaud_theory_2003}. Nevertheless, sampling from this long-tailed distribution remains a challenge, so preparing it as a quantum state and sampling qubit configurations from it presents itself as a potential practical application of quantum computers. In this work, we set the scale parameter to $c=32$.

To map continuous probability distributions onto a discrete set of quantum probability amplitudes, we use the \emph{quantics} representation \cite{oseledets_approximation_2009, khoromskij_quantics_2011}. First, we choose the interval $x \in [0,\ell)$ over which we want to encode the function, and then rescale as $\tilde{x}=x/\ell$, so that $\tilde{x} \in [0,1)$. For the Gaussian Eq.~\eqref{eq:gaussian}, we set $\ell=1$. For the L\'evy distribution Eq.~\eqref{eq:levy}, in order to capture the long tail over multiple lengthscales, we choose $\ell=2^{30}$. Then, assuming we are dealing with $N$ qubits, we discretize $\tilde{x}$ with a resolution of $1/2^N$,
\begin{equation}
	\tilde{x} = \sum_{i=1}^N s_i 2^{-i} \quad s_i \in \lbrace 0,1 \rbrace
\end{equation}
where $s_i$ are the qubit configurations, with the first qubit encoding the coarsest lengthscale of $2^{-1}$ (i.e., determining whether we are in the left or right part of the interval), down to the last qubit representing the finest lengthscale of $2^{-N}$. In this way, we can associate a qubit configuration $(s_1, s_2, \dots s_N)$ with a position $\tilde{x}$ in the interval with exponential accuracy $1/2^N$, and interpret the probability amplitudes as an $N$-dimensional tensor $F_{s_1 sa_2, \dots s_N}$. Then, we use the tensor cross interpolation to construct an MPS representation of the Gaussian on $N=30$ qubits, and the L\'evy distribution on $N=50$ qubits. Specifically, in the TCI of the Gaussian (L\'evy) distributions we performed $2$ ($3$) left-to-right and right-to-left sweeps, corresponding to $16020$ ($51244$) function evaluations, and converged to a solution with a maximal error of $\max_i \varepsilon_{i,i+1} = 10^{-12}$ (see Appendix~\ref{subsec:TCI}) and bond dimension $\chi=13$ in both cases. In this way we can construct accurate MPS representations of the considered functions without evaluating the functions on all $2^N$ arguments, which would be required for "standard" SVD-based tensor unfolding. After loading with TCI, the MPSs are normalized. 

\subsubsection{Lorenz attractor}
The next dataset comes from a multidimensional timeseries of the Lorenz system \cite{lorenz_deterministic_1963}, a classic example of a deterministic dynamical system leading to chaotic behavior. We generate the timeseries by integrating the equations of motion 
\begin{eqnarray}
	\frac{dx}{dt} &=& \sigma(y - x) \nonumber \\
	\frac{dy}{dt} &=& x(\rho - z) - y \nonumber \\
	\frac{dz}{dt} &=& xy - \beta z
\end{eqnarray}
with $\sigma=10$, $\rho=28$, $\beta=2.667$, starting from the initial condition $(x_0, y_0, z_0) = (0, 1, 1.05)$, for a total time $T=2^7=128$, with the Euler integration scheme and timestep $\Delta t=2^{-20}$. This choice of parameters makes the evolution concentrate around a butterfly-shaped chaotic attractor in the $x, y, z$ coordinates, see Fig.~\ref{fig:datasets} (third column). In total, we produce $T/\Delta t=2^{27}$ points for each of the $x,y,z$ axes. To perform the amplitude encoding, we transform the data into 1-dimensional time series of length $2^{29}$ by stacking the $x,y,z$ axes next to each other and padding the rest with zeros (adding a fictitious axis $w$ with zero coordinate at all times), so that the configuration of the first $2$ qubits determines the axis ($s_1s_2=00,01,10,11$ denote axes $x,y,z,w$), and the remaining $27$ qubits enumerate timesteps. The data is then normalized and transformed into an MPS on $N=29$ qubits with a step-by-step SVD with a relative singular value cutoff $\varepsilon_{\text{SVD}}=10^{-12}$, yielding a state with maximal bond dimension $\chi=92$. The transformation takes about $3$ minutes on a consumer-grade Apple Silicon M3 8-core CPU. 

\subsubsection{S\&P 500 stock prices}
Our last example is a publicly available financial dataset  Standard\&Poor's 500 (S\&P 500) \cite{nugent_sandp500_2018}, which contains timeseries of real stock prices of $505$ companies from the years 2013-2018. To encode it as a quantum state, we choose the first $1024=2^{10}$ day closing prices for each company and stack them next to each other, yielding a vector of length $505\times1024$, which we normalize to zero mean. Padding this vector with zeros by adding $7$ fictitious companies with zero price, we reach the size of $512\times 1024=2^{19}$, which we encode as probability amplitudes of a quantum state of $N=19$ qubits. The first $9$ qubits enumerate the companies, and the rest $10$ qubits enumerate points in time. The state is transformed into MPS form with the same method and parameters as for the Lorenz attractor dataset. The final MPS has a bond dimension of $\chi=512$, i.e. the maximal possible value for this system size.

\subsection{Entanglement structure analysis}
As it is demonstrated in Ref.~\cite{bohun_scalable_2025}, understanding the spatial entanglement structure of the states can be utilized to optimize their preparation as quantum circuits. Let us consider quantics encoding of function $f(\tilde{x})$, $\tilde{x}\in[0,1)$, as an MPS. A theorem of \cite{ripoll_quantum_2021} states that adding $(N+1)$-th qubit to a $N$-qubit state representing function $f(\tilde{x})$ can lead to a growth of the von Neumann entanglement entropy, but with an upper bound of
\begin{equation}
	\Delta \mathcal{S}_{\text{vN}} \leq \frac{\sqrt{\max\limits_{\tilde{x}\in[0,1]}\left|f'(\tilde{x})\right|}}{2^{N/2-1}} \sim \mathcal{O}(2^{-N/2})
\end{equation}
which decays exponentially with the number of qubits provided the function derivatives are finite. This suggests that smooth functions typically have weakly entangled quantic representations, and therefore are easy to transform into MPS e.g. through TCI, and to further prepare as quantum circuits. Ref.~\cite{bohun_scalable_2025} further studies the entanglement properties of smooth functions such as Gaussian and L\'evy distributions. If the function has a characteristic lengthscale $\xi$, below which there are no significant changes in the function value, one can distinguish two separate parts of the corresponding MPS. The first $j \approx \log_2 \xi$ bonds, between the qubits corresponding to the largest lengthscales, are characterized by a non-universal entanglement structure, which is dependent on the choice of the function $f$ and its large-scale variations. However, all further bonds $k>j$ are described by an universal exponential decay of the von Neumann entanglement entropy,
\begin{equation}
	\mathcal{S}_{\text{vN}}(k) \sim \mathcal{O}(\frac{j}{4^j}), \quad \text{as} \quad j \rightarrow \infty,
\end{equation}
for more details and formal definitions see \cite{bohun_scalable_2025}. This is not surprising - if the value of the function value stays the same irrespective of the state $0$ or $1$ of the $k$-th qubit encoding a small lengthscale $1/2^k \ll \xi$, this qubit carries no information, so there is no entanglement introduced with it.\footnote{The reverse is of course not true - if we interpret a random rank-1 MPS as a function $f(\tilde{x})$, the function will generically possess (highly structured) variations on the smallest lengthscale $1/2^N$ even though entanglement of the quantum state is trivially equal to zero everywhere.}

Fig.~\ref{fig:datasets} (middle row) shows the entanglement entropy at each bond for the four classical datasets. It is evident that in all cases there is an exponential decay of entanglement for the bonds with high indices. For the Gaussian, one expects $\xi = 2\sigma=0.2 \sim 1/2^2$ \cite{bohun_scalable_2025}, and indeed we observe an exponential decay after the first two qubits. For the L\'evy distribution, due to the power-law tail, there is no natural characteristic lengthscale; $\xi=2^{30}$ becomes the size of the support on which we consider the original function Eq.~\eqref{eq:levy}. For bonds $30$ to $50$ we observe an exponential decay entropy towards numerical zero, demonstrating that discretization of $\tilde{x}$ up to $\Delta \tilde{x}=1/2^{50}$ (correspondingly, $x$ to $\Delta x=1/2^{20}$) captures the behavior of $f(\tilde{x})$ exactly. The Lorenz attractor timeseries, except the "jumps" between parts of the data for different axes $x,y,z$, can also be treated as a smooth function and seems to have the same boundary around the $11$-th bond. The S\&P 500 stock prices are numbered by the first $9$ qubits, and the bonds in this area show a relatively large entanglement, $\mathcal{S}_{\text{vN}} \sim \mathcal{O}(1)$. There is still a slow exponential decay of entanglement on the remaining $10$ qubits encoding timesteps, suggesting that the temporal correlations present in the data allow for its efficient representation as MPS (the largest timescales of years are only weakly entangled with the smallest timescales - days).

Finally, Fig.~\ref{fig:datasets} (bottom row) shows the matrices of quantum mutual information $I_{ij}$ between all pairs of qubits. We find that in the Gaussian distribution case, most correlations in the system are restricted to the first $3$-$4$ qubits, and decay fast with distance. In the L\'evy distribution case, there are two disconnected clusters of strongly entangled qubits with a "regular" structure and the QMI decreases with the distance between qubits. The Lorenz attractor has a dominating (by an order of magnitude) QMI between qubits $1$ and $2$, which corresponds to the choice of the axis $x,y,z,w$ - this is fully expected as the fourth, "fictitious" $w$ axis, numbered by $s_1s_2=11$, has zero probability amplitudes which are very different from the remaining three configurations $s_1s_2=00,01,10$. This induces a strong correlation between qubits $1$ and $2$ - the quantum system must always satisfy $s_1=0$ whenever $s_2=1$ and vice versa. The remaining structure of QMI between different timescales is complex and the correlations spread in an all-to-all manner up to roughly $12$ qubits, above which the QMI decays to much smaller values. This happens due to the smoothness of the Lorenz attractor trajectory on a timescale below $\xi \approx 2^{29-12-2}\times\Delta t=2^{-5}$. In the S\&P 500 dataset, there are two clusters of entangled qubits in two approximately disconnected blocks of the matrix: qubits $1-9$, corresponding to the choice of the company, and qubits $10-19$, corresponding to the choice of the timestep. Clearly, the alphabetical initial ordering of the companies introduces a strong correlation between some pairs of qubits, visible as bright spots.

\subsection{MPS qubits reordering}
\label{subsec:results_reordering}

The MPS qubit reordering algorithm described in Sec.~\ref{subsec:reordering} is expected to bring qubits with a large QMI close together so that their entanglement does not need to be mediated by the MPS bonds inbetween. We apply this method to all datasets from Fig.~\ref{fig:datasets}. We find that in the Gaussian distribution case, since the QMI is concentrated only on the first 3-4 qubits, the original qubit arrangement is already optimal; no better solutions are found. The same is true for the L\'evy distribution data, for which the two clusters with a regular and seemingly optimal structure in the QMI are already formed in the standard quantics representation. 

In the Lorenz attractor case there is a group of approximately $10$ all-to-all entangled qubits with similar QMI, except for a special pair of strongly correlated qubits $1,2$, see discussion in Sec.~\ref{subsec:datasets}. The optimizer keeps this pair of qubits $1,2$ as neighbors, and moves the other qubits within the cluster, so that the cost function drops from $C=47.07$ to $C=45.25$, i.e. by less than $4\%$. Qualitatively, the structure of the QMI matrix remains the same (plot not shown). This is again not a good use case of the reordering algorithm because of the initially large all-to-all entanglement within the cluster, because after any reordering the cluster remains all-to-all correlated.

\begin{figure*}
    \includegraphics[width=\linewidth]{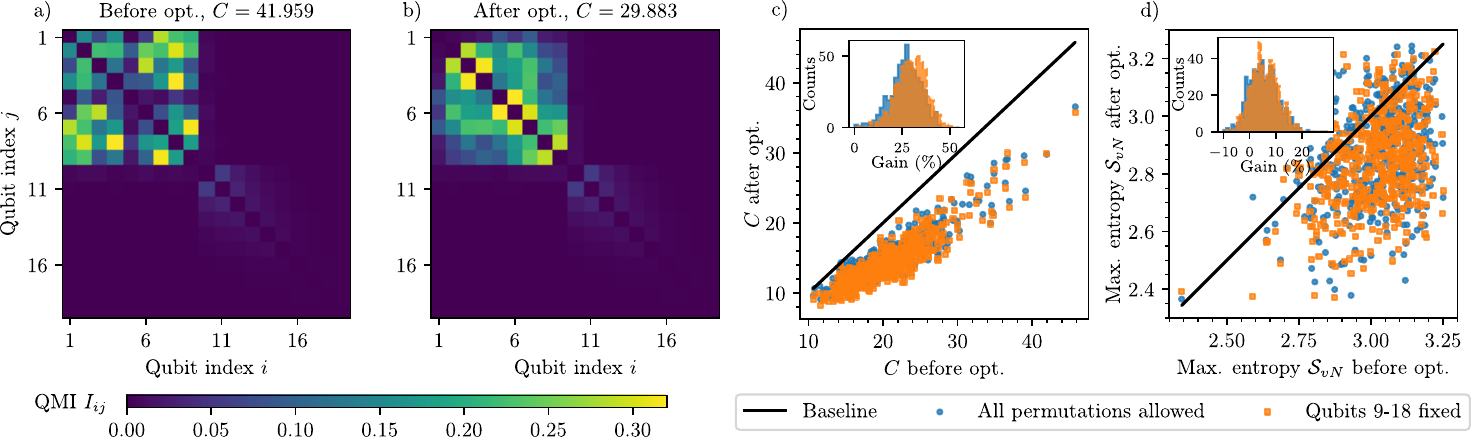}
    \caption{Qubit reordering problem solved as a QAP for the S\&P500 dataset. a) QMI between qubits for random realization of reshuffling the company indices. Qubits $1-9$, labeling the companies, show long-range correlations (off-diagonal peaks, upper-left rectangle).
    b) After the optimization, the cost $C$ decreases and the strongly correlated qubits are brought closer to each other.
    c) Optimized cost $C$ vs. pre-optimization cost for 500 random company permutations. Improvement is observed in nearly all cases. Colors indicate either unrestricted qubit permutations or fixed qubits 9-18 (timestep labels). Inset: histograms of relative cost improvement.
    d) Optimized maximal (over all MPS bonds) entanglement entropy $\SvN$ vs. pre-optimization maximal $\SvN$. In the majority of the cases, a decrease in the entanglement entropy is observed, even up to $20\%$, as shown in the inset.  
 }
\label{fig:sandp_reordered}
\end{figure*}

A successful application of the MPS qubit reordering algorithm is encountered for the S\&P 500 dataset. Since the QMI matrix has more pronounced high QMI isolated off-diagonal "peaks", we expect the reordering procedure to bring these peaks closer to the diagonal. To statistically verify the effectiveness of our method, we consider random permutations of an initially alphabetically ordered $505$ companies. One example of the QMI matrix resulting from a permutation of companies is presented in Fig.~\ref{fig:sandp_reordered}a), together with the result of an application of the reordering method to this state in Fig.~\ref{fig:sandp_reordered}b), with the allowed permutations restricted only to qubits $1-9$. This subset of permutations does not affect the ordering of qubits enumerating timesteps. Clearly, the algorithm simplifies the QMI matrix by moving the off-diagonal high-QMI peaks closer to the diagonal, yielding a $28\%$ decrease of the cost function \eqref{eq:cost}. In Fig.~\ref{fig:sandp_reordered}c) the initial and final cost function values are plotted for $500$ random permutations of the companies. We identify an improvement in the cost function in almost all realizations, on average reaching $30\%$. We also verify that the algorithm is indeed basing the solution of the problem on the permutation of the companies - if we allow to perform permutations on all qubits, "mixing" the company-enumerating and timestep-enumerating qubits, the average cost function improvement changes to $27\%$. It is smaller than for reordering restricted to qubits $1-9$ because the search space increases from $9!$ to $19!$, making the optimization harder. 

The final effect of qubit reordering can be measured e.g. by comparing the maximal von Neumann entanglement entropy on the original and reordered MPS bonds, see Fig.~\ref{fig:sandp_reordered}d). Indeed, for a majority of random realizations of the dataset (companies permutations), the maximal entanglement drops by a significant amount, in extreme cases even by $20\%$. The above results suggest that the qubit reordering algorithm can have a measurable effect on the bond entanglement whenever the QMI matrix is characterized by a "peaked", off-diagonal structure.

\subsection{Evaluation: SMPD/BMPD only}
\label{subsec:results_disentanglers}

\begin{figure*}[t!]
    \includegraphics[width=\linewidth]{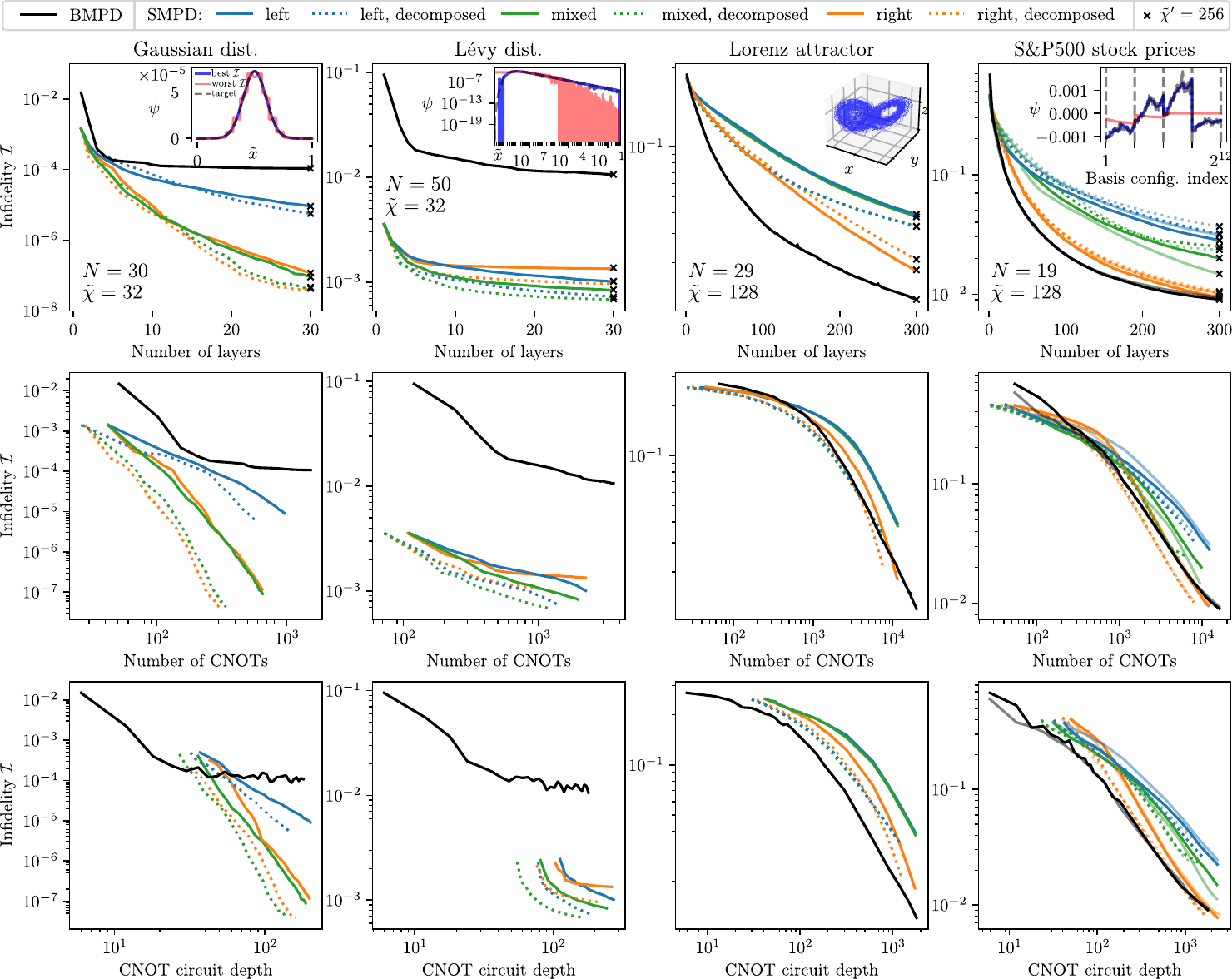}
    \caption{Infidelity as a function of number of disentangler layers (upper row), number of CNOT gates (middle row), and the CNOT circuit depth (bottom row) for different datasets (in columns). Black line represents the BMPD, while other lines represent the SMPD constructed upon left/right/mixed canonical form of the MPS (see Fig.~\ref{fig:full_pipeline_scheme}), as well as decomposition of isometries into two CNOT gates and single qubit unitaries (dotted lines) or no decomposition (2-qubit unitaries, solid lines). Insets show the real part of probability amplitudes, corresponding to preparation circuits with the worst (red, not shown for the Lorenz attractor) and best (blue) infidelity from the large plot, juxtaposed with the target amplitudes (black dashed line). For the S\&P 500 dataset, only the first $4$ companies ($4096$ amplitudes) are shown. Crosses represent an independent infidelity calculation with $\tilde{\chi}'=256$ after the circuit is fully constructed. For the S\&P 500 dataset, we also plot the values of infidelity after applying the MPS qubit reordering as "transparent" curves, see Sec.~\ref{subsec:results_reordering} for details.}
 \label{fig:results_disentanglers}
\end{figure*}

We start from the analysis and comparison of the baseline SMPD and BMPD algorithm without any subsequent gate optimizations.
Fig.~\ref{fig:results_disentanglers} shows the results of our calculations. Following the discussion in Sec.~\ref{subsec:smpd_convergence}, we choose the intermediate bond dimensions  $\tilde{\chi}$ as noted in the figures. We use the enhancement from Fig.~\ref{fig:enhancements_smpd}b) and cease to create SMPD or BMPD 2-body unitary gates around bonds that are of rank 1 at a relative SVD cutoff $\varepsilon_{\text{SVD}}=10^{-8}$ (Gaussian, L\'evy) or $\varepsilon_{\text{SVD}}=10^{-4}$ (Lorenz attractor, S\&P 500), based on the expected final fidelities computed separately without this enhancement. The orthogonality center position for the mixed canonical gauge SMPD is always $N/2$ to minimize the circuit depth, with an exception of the Gaussian for which we found it optimal to be at site $2$, see discussion in Appendix \ref{app:mixed_pos}. For the BMPD algorithm, we minimize the $\alpha$-R\'enyi entropy with $\alpha=2$ using the L-BFGS-B optimizer with at most $1000$ iterations. CNOT depth of the circuits is computed with \verb|qiskit v2.0.0| \cite{javadiabhari_quantum_2024}. The presented probability amplitudes are real - for presentation we remove the global phase based on the probability amplitude largest in magnitude and take the real part of the result. Calculations are performed with the help of \verb|quimb| \cite{gray_quimb_2018} library.

The upper row shows the state preparation infidelity $\mathcal{I}$ as a function of the number of layers of SMPD/BMPD\footnote{Note that the same symbol $L$ is used for the number of layers of the SMPD and BMPD, but a "layer" in both cases differ by gate arrangements and depth.}. It is clear that in all cases increasing the number of layers decreases the infidelity. In the Gaussian case, already a single SMPD (BMPD) layer gives a reasonably good approximation of the state with infidelity $\Ical=10^{-3}(10^{-2})$, the latter shown as a red curve in the inset. We observe that the BMPD algorithm saturates at around $\Ical=10^{-4}$ for $L\gtrapprox 10$ layers, whereas the SMPD algorithm converges to a perfectly prepared state with $\Ical<10^{-7}$ for $L=30$ layers (blue curve in the inset). Convergence of the MPS simulation is checked independently by computing the infidelity with bond dimension $\tilde{\chi}'=64,128,256$ (only $\tilde{\chi}'=256$ shown) after the QSP circuit is fully constructed with an intermediate $\tilde{\chi}=32$, see black crosses in Fig.~\ref{fig:results_disentanglers} which match the $\tilde{\chi}=32$ values. It is also evident that one has to be careful when choosing the canonical form of the MPS for SMPD - changing from the left canonical form to the right or mixed canonical form for this dataset improves the final infidelity by $2$ orders of magnitude. The imbalance is caused by the concentration of entanglement around the first $3$-$4$ sites, see Fig.~\ref{fig:datasets}, and the opposite order of gate application in the left/right SMPD. Finally, the (analytical) decomposition of the two-body isometric gates into the form presented in Fig.~\ref{fig:enhancements_smpd}a), i.e., with $2$ CNOT gates instead of $3$ CNOT gates, does not negatively affect the overall state preparation efficiency. In fact, for the Gaussian distribution we observe a slight improvement of infidelity after turning on this enhancement, at the same time saving on $33\%$ CNOT gate number.  Middle and bottom rows show the infidelity as a function of the number of CNOTs $\ncnot$ and the CNOT circuit depth $\dcnot$. The right canonical SMPD with decomposed gates prepares the state with the smallest $\ncnot$ at a given fidelity. The mixed canonical SMPD with decomposition reaches comparable infidelities for the same number of CNOTs, but has a slightly smaller CNOT depth. Moreover, due to the architecture, a few layers of BMPD can prepare the state with a much lower $\dcnot$ than even a single layer of SMPD, but SMPD becomes more effective when depth exceeds $\dcnot \gtrapprox 30$.

For the L\'evy distribution, our calculations show qualitatively similar trends. Here, the SMPD yields an order of magnitude better infidelity than BMPD for a comparable $\ncnot$ and $\dcnot$. With SMPD we do not reach infidelities better than $\Ical \approx 6\times 10^{-4}$ for $L=30$ layers. As shown in the inset, the areas of the $\tilde{x}$ axis where the probability amplitude is lower than $10^{-7}$, for $\tilde{x}\lessapprox 10^{-9}$ are not reconstructed faithfully, but this is expected, as their absolute contribution to the fidelity is very small, $\mathcal{O}(10^{-14})$. On the contrary, the heavy tail is correctly reconstructed. This time, the left, mixed, and right canonical forms of the MPS do not yield infidelities differing by more than $\Delta \Ical=0.002$.

The remaining two examples, due to a much richer entanglement structure, see Fig.~\ref{fig:datasets}, required deeper circuits of $L=300$ layers to reach $\Ical \approx 10^{-2}$. This value lets us visually distinguish the features of the prepared and shown Lorenz attractor, as well as a faithfully reproduce the prices in time for the S\&P 500 dataset, see insets for both. Importantly, we demonstrate that in both cases the infidelities slowly but steadily improve for $L=300$ layers, also for circuits with decomposed 1-to-2 qubit isometries. Unlike for the previous two simpler states, here the BMPD yields a slightly better preparation infidelity than any variant of SMPD for the same $\dcnot$, except for the shallowest circuits which are not achievable with SMPDs, and comparable infidelities for the same $\ncnot$. Note that we were able to prepare both of these states with $\tilde{\chi}=128$, even though their corresponding bond dimensions are $\chi=92$ and $\chi=512$, respectively. For the S\&P 500 dataset, this introduces a minimal value of the infidelity of $\Ical=6.8\times 10^{-5}$, resulting from its truncation to the closest $\chi=128$ state. We checked that the final infidelity does not change when calculated with $\tilde{\chi}'=512$. 

Another important observation is that the improvements of infidelity after the application of the MPS qubit reordering are very minor, see the "transparent" curves for the S\&P 500 dataset in Fig.~\ref{fig:results_disentanglers}. We find this rather surprising, as the qubits with the strongest correlations become nearest neighbors. Nevertheless, these results do not exclude potential gains after applying the reordering procedure to different target states, such as the groundstates of chemical molecules in quantum chemistry, for which computational gains after reordering have been confirmed \cite{barcza_quantum-information_2011, rissler_measuring_2006}.

Differences of infidelity between the SMPD variants with decomposed and not decomposed gates for the same other settings suggest an interesting avenue for future research. Since the decomposition of the isometry is in fact an alternative way to complete the isometric matrix to a unitary matrix by adding orthogonal columns, one could investigate different such methods, potentially taking into account the effect on the next layer, and find the one best suited for the problem.

\subsection{Evaluation: SMPD/BMPD with optimization}
\label{subsec:results_optimized}
In Sec.~\ref{subsec:results_disentanglers} we have described the performance of SMPD and BMPD as a state preparation method. Although we have shown that both methods yield reasonably good solutions under moderate classical numerical cost for each dataset, we either run into a problem of infidelity saturating with the number of layers due to the heurstic nature of both methods, or an exceedingly high and impractical gate counts/depths. In this section we improve on the QSP methods from Sec.~\ref{subsec:results_disentanglers} by extending them with the Evenbly-Vidal (EV) optimization, as well as with the Riemannian (R) optimization. 

Inspired by the results of \cite{rudolph_decomposition_2023}, we follow two strategies in combining the two methods into a state preparation pipeline. The first strategy is to simply create a $L$-layer disentangler SMPD/BMPD and then optimize the unitary gates with the EV or Riemannian optimizer (referred to as SMPD-EV, SMPD-R, BMPD-EV, BMPD-R accordingly). The second strategy \cite{rudolph_decomposition_2023} is to create a single layer of SMPD/BMPD, optimize all gates with EV or Riemannian optimizer, create another layer with SMPD/BMPD, optimize all gates, and so on (referred to as SMPD-EV-SMPD-EV, SMPD-R-SMPD-R, etc.). For an improved readability, the results for the SMPD and BMPD optimization are analyzed separately in Figs.~\ref{fig:smpd_optimized},~\ref{fig:bmpd_optimized}. Their performance with respect to $\ncnot$ and $\dcnot$ is compared in Fig.~\ref{fig:both_optimized}, where they are shown together. For each benchmarked configuration, we assume a maximal computational budget of $24$ hours on $4$ to $6$ AMD Epyc 7443 CPU cores and $24$ GB DDR4 RAM memory, i.e. we terminate the calculation after the time or memory limit is reached. For the Riemannian Adam optimizer we choose an initial learning rate $lr=10^{-4}$ and perform $\niterR=10^4$ iterations for the SMPD-R/BMPD-R case and $\niterR=10^3$ iterations at each of $L$ steps of the SMPD-R-SMPD-R/BMPD-R-BMPD-R case. In the EV optimization, we use the learning rate of $\beta=0.6$, $\nswpEV=10^2$ sweeps for the SMPD-EV-SMPD-EV/BMPD-EV-BMPD-EV case and $\nswpEV=10^3$ sweeps for the SMPD-EV/BMPD-EV case. Before the EV/R optimization, we identify pairs of neighboring single-body and two-body gates on the same qubits, whose action can be absorbed into a single- or two-qubit gate. This applies whenever there are no other gates between them acting on the same qubits and can be understood as a simple and deterministic transpilation step that removes redundant operations without changing the final state produced by the circuit. In this way, we improve the performance of the EV/R optimizers by reducing the number of variational parameters, including the linearly dependent ones.

\subsubsection{SMPD} 
    \begin{figure*}[t!]
	\includegraphics[width=\linewidth]{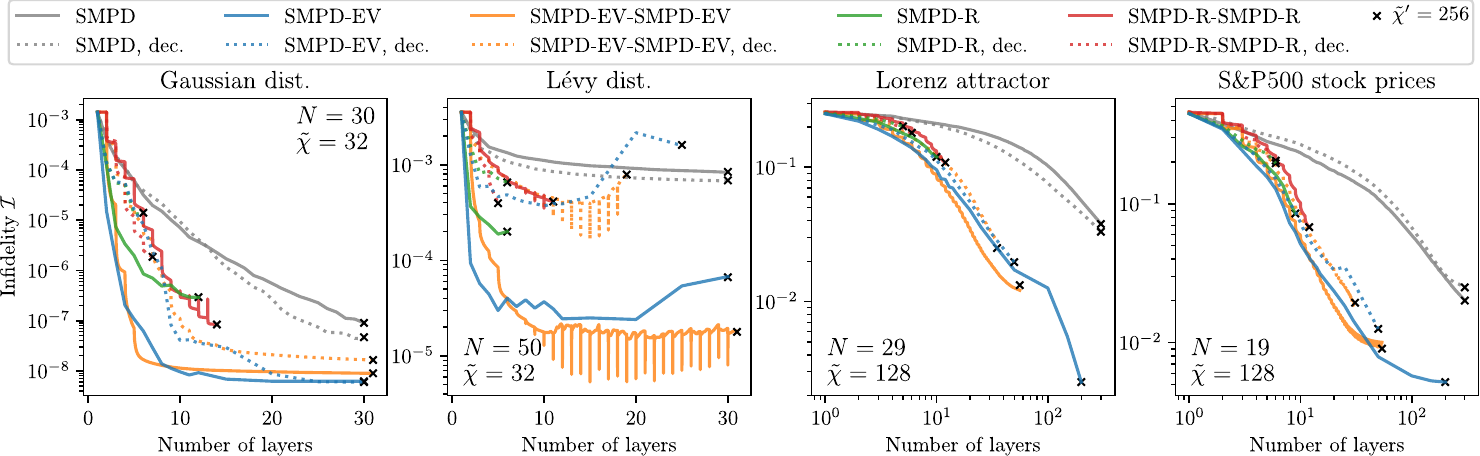}
	\caption{State preparation infidelity of SMPD circuits combined with Evenbly-Vidal or Riemannian optimization. Black crosses represent $\tilde{\chi}'=256$ verification after circuit construction. Intermediate values between layers show infidelities at each optimization step (orange, red), or join neighboring points by a straight line for visibility (gray, blue, green). Isometry decomposition (dotted lines) gives a number of CNOT and CNOT depth benefit for shallow circuits, depending on the dataset.}
	\label{fig:smpd_optimized}
\end{figure*}

Fig.~\ref{fig:smpd_optimized} shows the results of the SMPD circuits optimization. We have selected the mixed canonical gauge SMPDs from Fig.~\ref{fig:results_disentanglers} to minimize the total depth of the circuit without sacrificing too much fidelity, both with and without isometry decomposition (gray lines). For optimization, the gates are kept as 2-qubit $4\times 4$ unitary matrices in the standard case\footnote{If a certain bond has already been disentangled by the SMPD to a sufficient level, the gate in SMPD becomes a single-body unitary; it is also treated as variational gate.}, and as a set of single-qubit $2\times 2$ variational unitaries sandwiched between fixed CNOTs in the decomposed case. Note that this means a smaller number of free parameters per gate ($4$ vs $16$ complex numbers) but roughly six times more variational gates in the decomposed case. Rotation gates such as $R_x$, $R_z$ are "upgraded" to general unitaries. Between layers $L$, $L+1$ we plot the trajectory of the infidelity in the optimizations (SMPD-R-SMPD-R, SMPD-EV-SMPD-EV, etc.), or join the infidelity points by a straight line (SMPD-R, SMPD-EV).

 \begin{figure*}[t!]
	\includegraphics[width=\linewidth]{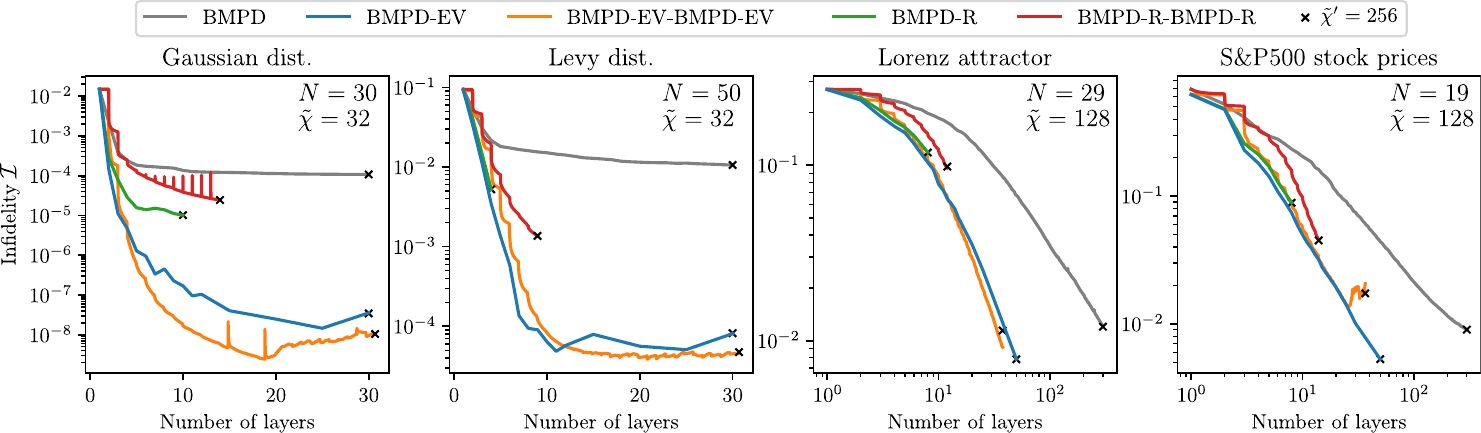}
	\caption{State preparation infidelity for the BMPD method (gray, baseline) combined with the EV/R optimization methods. Convergence is verified by computing infidelities with a higher bond dimension $\tilde{\chi}=256$. Clearly, applying any optimization method leads to orders-of-magnitude decrease of infidelities in all cases. The BMPD-EV and BMPD-EV-BMPD-EV methods give the best infidelity for a given number of BMPD layers.}
	\label{fig:bmpd_optimized}
\end{figure*}

In the top row of Fig.~\ref{fig:smpd_optimized} we present the state preparation infidelity as a function of the number of layers. We ensure convergence with the bond dimension $\tilde{\chi}$- all final infidelities computed with an intermediate bond dimension $\tilde{\chi}=32$ or $\tilde{\chi}=128$ match with a later verification of the same circuits for $\tilde{\chi}'=256$ (black crosses). 
Note that Riemannian optimization is not affected by this setting since the tensor network is contracted exactly. For all optimization configurations we observe an abrupt drop of infidelity starting from $L=1$ when compared to the baseline. However, there are orders-of-magnitude differences between methods depending on the dataset. In all cases, the SMPD-EV and SMPD-EV-SMPD-EV methods seem to be the most efficient. Moreover, using the decomposition is favorable only for the shallowest circuits of a few layers, with the number depending on the dataset, when the advantage of a roughly 33\% CNOT count and depth outweighs potential problems with optimization of more single-qubit $2\times2$ unitaries instead of less two-qubit $4\times 4$ unitaries. 

Due to the global nature of the Riemannian optimizer, updating all unitary gates at once, we expected it to outperform the EV optimization sweeping over individual gates \cite{luchnikov_qgopt_2021, hauru_riemannian_2021, melnikov_quantum_2023}. At least for the four considered datasets we conclude that the EV sweeping optimization is generally more performant in terms of state preparation fidelity. The EV method is also more scalable because it keeps the circuit state in the MPS form throughout the calculation, so that we can reach up to $L=200$ optimized layers in the Lorenz attractor and the S\&P $500$ case (note, however, that no strict bounds on $\tilde{\chi}$ can be given in general). Since the Riemannian optimizer deals with exact contractions of tensor networks with unitary gates as elements, we cannot go beyond $L\approx \mathcal{O}(10-15)$ layers for the considered system sizes, because the sizes of intermediate tensors exceed our memory limitation of $24$ GB and grow exponentially with the number of layers. We do not exclude that using higher-order optimizers for the Riemannian method, instead of Riemannian Adam, could lead to improvement in convergence, because Hessian could capture more information on the interactions between elements in different unitary matrices, see e.g. Ref.~\cite{hauru_riemannian_2021} for L-BFGS on Riemannian manifolds in the groundstate search context. Moreover, to deal with deep circuits and the Riemannian optimization, one could use a divide-and-conquer approach: optimize only $\mathcal{O}(10-15)$ layers at once, and keep the intermediate states to the right and to the left of these gates in the MPS form. We leave tests of this variation of the method to future works. 

We notice problems with optimization in the L\'evy distribution SMPD-EV-SMPD-EV case. After $L=10$ layers are reached, the infidelity "jumps" down after adding a new layer of SMPD, but increases back during the EV optimization. 
Decreasing the learning rate $\beta$ of the EV optimizer does not resolve the issue. We find that this behavior is caused by a too small intermediate bond dimension $\tilde{\chi}=32$. Even though the final infidelity computed with $\tilde{\chi}'=256$ matches with the $\tilde{\chi}=32$, so the result can be regarded as converged, the optimizer needed a larger intermediate $\tilde{\chi}$ to explore a larger space on the way to solution. When computed with a higher $\tilde{\chi}=128$, the "jumps" are much smaller and start to appear later, around $L=18$. This leaves us with a useful additional diagnostic to determine what size of $\tilde{\chi}$ is required.

We observe a similar "spiking" behavior (but of different origin) in the SMPD-R-SMPD-R case for the Gaussian distribution with an initial learning rate of $lr=10^{-4}$. After setting $lr=10^{-5}$ (shown in the plot), the "spikes" in infidelity disappear and the solution converges to a better final infidelity. 

For the Lorenz attractor and S\&P 500 dataset, represented by highly entangled states, we observe that all optimization methods perform comparably well, approximately forming a universal curve in the $\Ical(L)$ plot. This would suggest that for such states one is left with a choice between the smallest computational cost, in this case SMPD-EV, and the best infidelity for the same number of CNOTs and CNOT depth, here SMPD-EV supplemented with isometry decompositions.

\subsubsection{BMPD}

 \begin{figure*}[t!]
	\includegraphics[width=\linewidth]{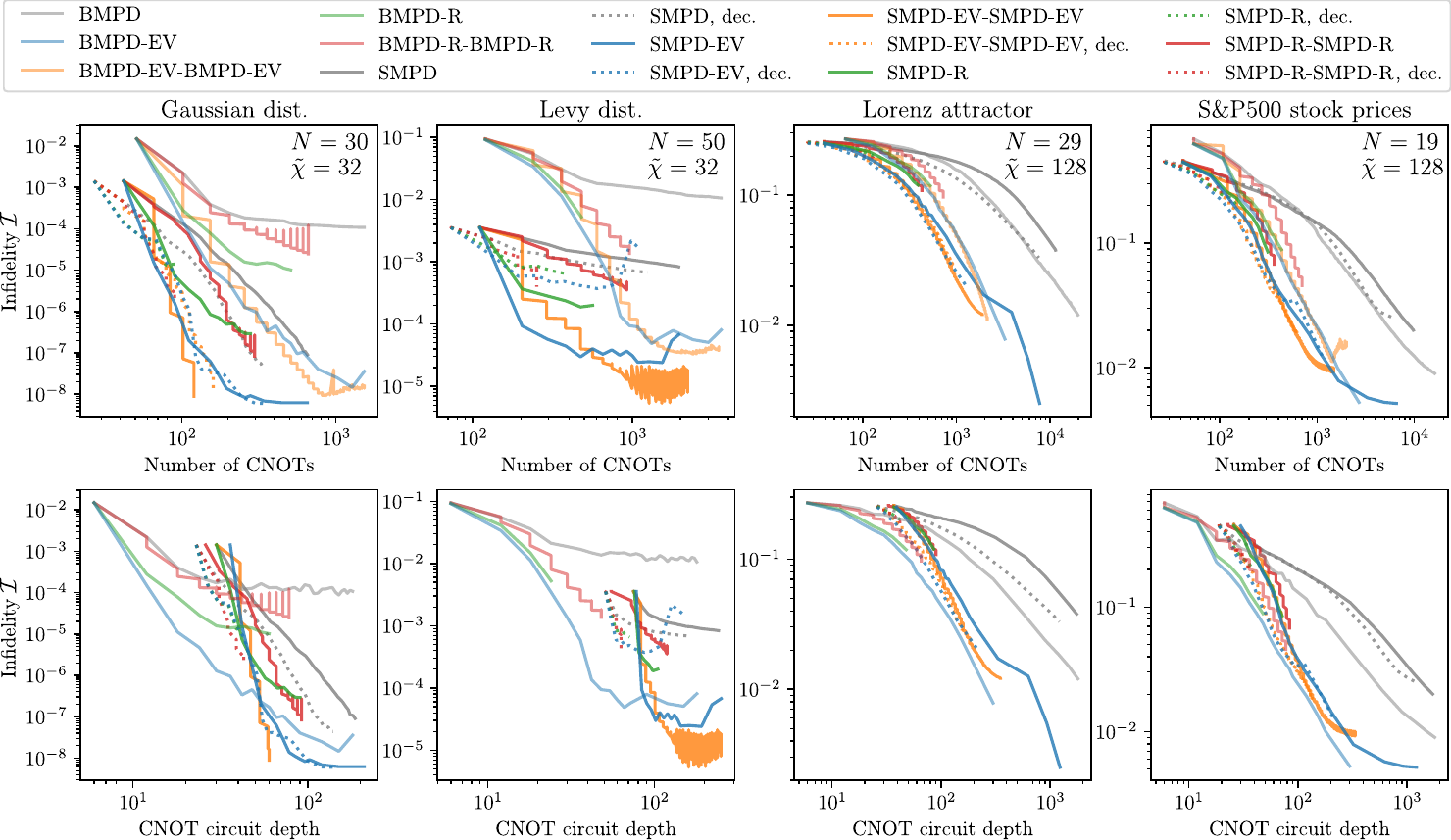}
	\caption{Comparison of infidelities obtained with the BMPD (light colors) and SMPD (regular colors) circuits supplemented with EV and Riemannian optimization methods. If the CNOT gate count is a priority, typically the SMPD-EV and SMPD-EV-SMPD-EV methods are a good choice. To minimize the CNOT depth, one should resort to BMPD-EV and BMPD-EV-BMPD-EV methods, especially for the shallowest circuits made of a few disentangler layers.}
	\label{fig:both_optimized}
\end{figure*}
Analogously to the above SMPD case, we perform a study of the combination of BMPD with EV/R optimization techniques. Results are presented in Fig.~\ref{fig:bmpd_optimized}. Again, the results for the Riemannian optimizers cannot be obtained for more than $L\approx \mathcal{O}(10)$ layers due to an increasing size of intermediate tensors in the tensor network contraction. The infidelities converge with the bond dimension $\tilde{\chi}$, as we confirm by computing them with $\tilde{\chi}'=256$ for the circuits obtained with $\tilde{\chi}$. 

In general, in Fig.~\ref{fig:bmpd_optimized} we observe orders-of-magnitude better state preparation infidelities than with the bare BMPD method. Each gate in the heuristic BMPD circuit is constructed in such a way that it disentangles a single bond. When further optimization is applied, an update for a single gate results from its interplay also with other gates and bonds. Thus, by "cooperation" between gates, one can find the global solution with much less layers and to a higher fidelity. Nevertheless, our understanding is that a heuristic initialization with BMPD is indispensable, as it sets up an initial point not only in terms of a non-vanishing initial fidelity, giving non-zero gradients of the gate parameters at the initial stages of optimization, but also in terms of a "regular" layer-by-layer path to the solution through intermediate states with a smoothly decreasing entanglement, see e.g.~\cite{mansuroglu_preparation_2025}(Fig.~3), which is refined by the optimizer.

We find that for a small number of layers, $L<5$, the best two methods are BMPD-EV and BMPD-R, whose infidelity curves are barely distinguishable. The benefit of interleaved BMPD-EV-BMPD-EV starts to appear later, with this method overtaking BMPD-EV for the deepest circuits, although maximally by one order of magnitude of infidelity $\Ical$. Given a larger numerical cost of BMPD-EV-BMPD-EV, the default choice for the state preparation would be BMPD-EV.

\subsubsection{Comparison}

Practical conclusions can be drawn from a comparison of the SMPD and BMPD supplemented with optimizers in terms of infidelities reached for the same CNOT count or CNOT depth, see Fig.~\ref{fig:both_optimized}. We do not compare the performance for the same number of layers, as the SMPD and BMPD have different circuit architectures. We notice that the SMPD method almost always outperforms BMPD in terms of the CNOT count required to reach the same infidelity across all infidelity ranges (upper row). On the other hand, the BMPD circuits are advantageous whenever the CNOT depth of the circuit is a priority (bottom row). This occurs because the depth of SMPD is $\approx \mathcal{O}(N/2+L)$, and for small $L \ll N$ and $N\gg 1$ it becomes larger than $\approx \mathcal{O}(2L)$ for BMPD.  This suggests that for each practical realization with noisy circuits, the tradeoff between the circuit depth and gate count has to be taken into account on a case-by-case basis.

\section{Conclusions}
\label{sec:conclusions}

This work introduces an MPS-based pipeline for quantum state preparation (Fig.~\ref{fig:full_pipeline_scheme}). The pipeline combines sequential and brick-wall disentangler circuits to generate an approximate initial state, which is subsequently refined using variational Evenbly–Vidal and Riemannian optimization techniques. In addition, we incorporate technical improvements aimed at reducing CNOT gate counts and circuit depth within a unified framework. A systematic study on systems of $N=19$–$50$ qubits demonstrates how the optimal choice of methods depends on the entanglement structure of the dataset and on the targeted performance metric, such as infidelity, gate count, or circuit depth.

Starting from the bare SMPD/BMPD algorithms, we find that for simple states (Gaussian and L\'evy distributions) the SMPD method outperforms BMPD in all aspects (CNOT depth and count for a given infidelity) except for the CNOT depth in the shallowest circuits. For more complex states (Lorenz attractor, S\&P 500) we need $L=300$ layers to achieve fidelities close to $\Fcal=0.99$. In this case, BMPD gives circuits with a consistently smaller depth. We also identify a strong interplay between the entanglement structure and the MPS left/right/mixed canonical form chosen to construct the SMPD. A clever choice of the gauge for the specific state can lead to 2 orders of magnitude improvement of infidelity $\Ical$, as we observe for the Gaussian distribution.  
1-to-2 qubit isometry decomposition does not spoil the performance of SMPD while decreasing the CNOT count and depth by 33\%. We prepare the four states with infidelities ranging from $\Ical=10^{-7}$ to $\Ical=10^{-2}$, but we also identify either large circuit depth requirement, or a saturation/slow convergence of the heuristic SMPD/BMPD methods with an increasing number of layers. Therefore, variational optimizers are employed to push for more efficient circuits.

As expected, the use of EV/R optimizers on precomputed SMPD/BMPD circuit Ans\"atze leads to a significantly faster convergence of infidelity with the number of layers. Quite surprisingly, the EV sweeping optimization approach seems to outperform the global Riemannian optimizer in terms of the final infidelity for the same number of BMPD and SMPD layers. Due to the MPS form retained throughout the EV optimization, one can also optimize deep circuits (we demonstrate $L=200$ layers of SMPD with $\ncnot\sim \mathcal{O}(2\times 10^3)$, $\dcnot\sim \mathcal{O}(10^3)$ for $N=29$ qubits for the Lorenz attractor), whereas the Riemannian optimization is restricted to relatively shallow circuits ($\ncnot\sim \mathcal{O}(5\times10^2)$, $\dcnot \sim \mathcal{O}(10^2)$, or $L\sim \mathcal{O}(10-15)$ layers), which is caused by exponentially growing sizes of intermediate tensors in the tensor network contractions. Variational gate optimization can be combined with isometry decompositions saving 33\% CNOTs, yielding better solutions especially for the shallowest circuits. However, for $L\gg1$ layers, it negatively affects the optimization for simple target states.

Finally, we identify the globally most performant combination of the presented methods across our datasets. We find that the SMPD-EV and SMPD-EV-SMPD-EV, potentially supplemented with isometry decompositons, are the optimal choices if the minimal CNOT gate count is a priority. For a minimal CNOT depth, one should rather resort to the BMPD-EV or BMPD-EV-BMPD-EV methods. 

We also provide an independent analysis of the SMPD classical simulability. Our results suggest that the minimal bond dimension $\tilde{\chi}$ sufficient to construct $L$ SMPD layers is on the order of the target MPS bond dimension, $\tilde{\chi}\sim \mathcal{O}(\chi)$, in contrast to the exponential complexity $\tilde{\chi}\sim\mathcal{O}(2^L)$ postulated in Ref.~\cite{ran_encoding_2020}. 

The pipeline also incorporates an entanglement-based qubit reordering scheme. Our contribution here is a mapping of this task into an instance of the quadratic assignment problem. We find that the qubit reordering algorithm has a rather limited applicability for the first three benchmarked states. While the Gaussian and L\'evy distribution MPSs already have their qubits arranged in the optimal order, the Lorenz attractor dataset shows significant and uniform all-to-all entanglement among a group of $12$ neighboring qubits, which does not allow us to decrease bond entanglement. The stock market companies from the S\&P 500 dataset, which give a "peaked" structure in the QMI matrix of the MPS, leads to a successful decrease of the maximal bond entanglement by up to 20\%. However, we do not observe any significant improvements in state preparation, but expect them for other classes of target states such as groundstates of molecules \cite{barcza_quantum-information_2011, rissler_measuring_2006}.

Let us briefly identify possible extensions of our work. The described QSP algorithms for MPS can be readily used for pretraining variational quantum circuits with classical computers \cite{rudolph_synergistic_2023}, which is also intimately related to approximate quantum compilation from \cite{jaderberg_variational_2025}. As already mentioned in Sec.~\ref{subsec:results_optimized}, the Riemannian optimization could potentially benefit from a second-order optimizer such as L-BFGS-B \cite{hauru_riemannian_2021, hosseini_alternative_2020}. Moreover, we believe it would be an interesting path of future research to include noise into the optimization pipeline which could potentially lead to better fidelities of real hardware realizations.

\paragraph*{Acknowledgements}
P. S. and T. S. acknowledge the funding of the Federal Ministry of Research, Technology and Space Project Number 13N17157 (Q-ROM). R. M. acknowledges support from the U.S. National Institute of Standards and Technology (NIST) through the CIPP program under Award No. 60NANB24D218. and the U.S. National Science Foundation (NSF) through the NSF TIP program under Award No. 2534232. The computational resources were provided by the PHYSnet-Rechenzentrum of Universität Hamburg and we thank M. Stieben for technical support. T. S. thanks Z. Zeybek for helpful discussions. T.S. and P.S. acknowledge inspiring conversations with C. Blank and I. F. Araujo on the use of QMI graphs in quantum state preparation.


\setcounter{equation}{0}
\setcounter{figure}{0}
\setcounter{table}{0}
\makeatletter
\renewcommand{\theequation}{A\arabic{equation}}
\renewcommand{\thefigure}{A\arabic{figure}}
\renewcommand{\thetable}{A\arabic{table}}

\onecolumn
\appendix

\section{Tensor Cross Interpolation}
\label{subsec:TCI}
Consider an $N$-dimensional tensor $F_{s_1 s_2 \dots s_N}$ with discrete indices $s_1, s_2,...,s_N$, each of dimension $d$ (in our case, $d=2$ for qubits). If the tensor has a low-rank structure, its decomposition into a matrix product state form, for example by a sequence of truncated SVDs or rank-revealing QR or LU decompositions, can give its more storage- and compute-efficient representation. However, to perform the SVDs, one first needs to access all $d^N$ elements of $F_{s_1 s_2 \dots s_N}$, even if the final result is an MPS of a very low bond dimension and contains much less parameters than $d^N$. This redundancy suggests that if the tensor $F_{s_1 s_2 \dots s_N}$ has a low-rank structure, it should be possible to construct its MPS approximation without calling it on all of its $d^N$ indices, which is the goal of Tensor Cross Interpolation (TCI). TCI treats the tensor as a black-box function and calls it only on $\mathcal{O}(Nd\chi^2)$ arguments. This section is a high-level introduction to TCI. For a more rigorous and complete exposition, we refer the reader to Refs.~\cite[Sec. III]{nunez_fernandez_learning_2022}, \cite{dolgov_parallel_2020}, or the more recent Refs.~\cite{fernandez_learning_2025,waintal_who_2026}. We start from the simpler case of low-rank approximations of matrices without accessing all of their elements. We then generalize to multidimensional tensors.

\subsection{Matrices}
Consider a real or complex matrix $A$ of dimension $n \times m$. In general, it requires a storage of $n\cdot m$ elements. However, if a certain structure is present in the matrix, which is typical for matrices carrying some information except for a random noise, its low-rank approximation may be sufficient to faithfully represent it with a smaller number of elements. A widely adopted algorithm for this task is the truncated SVD, which gives an optimal, in the Frobenius norm sense, approximation at a given rank $\chi$ by keeping $\chi$ largest singular values and setting the rest to zero. However, the SVD requires the knowledge of \emph{all} matrix elements $A(i,j)$, whose computation and storage may become a bottleneck for very large matrices.

An alternative way of rank-$\chi$ factorization is by using only a subset of the matrix elements: $\chi$ rows $\Ical = \left\lbrace \Ical^{(s)}\right\rbrace_{s=1}^\chi$ and $\chi$ columns $\Jcal = \left\lbrace \Jcal^{(t)}\right\rbrace_{t=1}^\chi$,
\begin{equation}
	A(i,j) \approx \tilde{A}(i,j) = \sum_{t,s=1}^\chi A(i, \Jcal^{(t)})
	\left[ A(\Ical,\Jcal)\right]_{t,s}^{-1} A(\Ical^{(s)},j),
	\label{eq:CI}
\end{equation}
or graphically,
\begin{center}
\includegraphics[width=.5\linewidth]{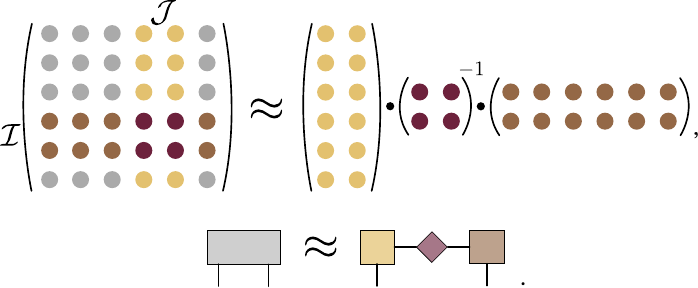}
\end{center}
which is known as \emph{cross interpolation} (CI) \cite{goreinov_pseudoskeleton_1997}. To compute the right hand side, we only need columns $A(i, \Jcal^{(t)})$ and rows $A(\Ical^{(s)},j)$, without accessing the rest of the matrix. Elements of the nonsingular submatrix $A(\Ical, \Jcal)$ of $A$ are called \emph{pivots}, and they form a \emph{pivot matrix}. Eq.~\eqref{eq:CI} is indeed an \emph{interpolation} of matrix $A$ on its selected columns and rows; one can show that by definition it exactly reproduces $A(i,\cdot)$ for any $i\in\Ical$ and $A(\cdot, j)$ for any $j\in\Jcal$. The name \emph{cross interpolation} (CI) comes from the "cross" shape of the submatrices of $A$, as depicted above. The accuracy of the approximation depends on the number and choice of the pivots, and can be systematically improved. Unlike the SVD, the method is not optimal, but quasioptimal in the approximation error which is $\mathcal{O}(\chi^2)$ times the optimal error \cite{savostyanov_quasioptimality_2014}. When $\text{rank}(A)$ is equal to the number of pivots $\chi$, one can obtain an exact factorization, $\tilde{A}=A$.

The main difficulty in cross interpolation lies in a proper selection of pivots, which should minimize the approximation error. In principle, there are exponentially many possible choices, and a brute-force check of all of them becomes infeasible even for modestly sized matrices. There exist successful heuristic approaches which, given the matrix $A$ and the current pivot matrix $A(\Ical, \Jcal)$, propose a new pivot $A(x_0, y_0)$, with the aim of maximizing the magnitude of the determinant of the new pivot matrix, $|\det A(\Ical\cup \lbrace x_0 \rbrace, \Jcal \cup \lbrace y_0 \rbrace )|$. This is also known as the \emph{maxvol} or Maximum Volume principle \cite{goreinov_pseudoskeleton_1997, goreinov_how_2010}. It turns out that by maximizing the magnitude of the determinant, one simultaneously chooses the row $x_0$ and the column $y_0$ of $A$ for which the current approximation error $\varepsilon_{x_0,y_0} = |A(x_0,y_0)-\tilde{A}(x_0,y_0)|$ is maximal, i.e., which contains the most "valuable" information that leads to the largest improvement in the total error (once $A(x_0,y_0)$ is added to the pivot matrix, it will be interpolated exactly with zero error), for a proof see \cite{nunez_fernandez_learning_2022}(Appendix B.2). Determination of the new pivot can be done with $\mathcal{O}(m+n)$ function calls with a heuristic "rook" strategy: from a random starting point $(i_0,j_0) \notin (\Ical, \Jcal)$ one scans over all rows $i=1\dots m$ of $A$ to find the one $i^*$ with the largest error $\varepsilon_{i^*,j_0} = |A(i^*, j_0)-\tilde{A}(i^*,j_0)|$ with $\mathcal{O}(m)$ function calls, and then scans over all columns $j=1\dots n$ to maximize an analogous $\varepsilon_{i^*,j^*}$ by choosing $j^*$ with $\mathcal{O}(n)$ function calls. Iteration between choosing rows and columns is repeated a few times, typically $3-5$, or until the pivot does not change \cite{dolgov_parallel_2020}. A complete CI of a matrix of rank $\chi$ requires $\chi$ pivot points, hence in total one needs to perform $\mathcal{O}(\chi(n+m))$ function calls.

\subsection{Tensors}

Like with SVD, one can decompose the tensor $F_{s_1 s_2 \dots s_N}$ into an MPS form, site-by-site reshaping it into a matrix, performing the matrix cross interpolation, and sweeping throughout all site tensors \cite{oseledets_tt-cross_2010}. Since every site tensor in this case is a result of an interpolation, exact on the chosen set of pivots, the whole MPS approximation is itself exact on the pivots used to construct it. To get MPS structure without the inverse pivot matrices between sites, one can contract them with site tensors. Although this approach provides a conceptual link between the matrix cross interpolation and the tensor cross interpolation (TCI), this single-site update approach seems to be ineffective in practice \cite{dolgov_parallel_2020}. 

A more effective algorithm \cite{dolgov_parallel_2020} is based on "superblock" tensors $\Pi_{i,i+1}=A_i^{(s_i)}A_{i+1}^{(s_{i+1})}$, which are a contraction of two neighboring MPS site tensors. In one variant of the algorithm one directly cross-interpolates the $\Pi_{i,i+1}$ tensor by reshaping it into a matrix \cite{dolgov_parallel_2020, nunez_fernandez_learning_2022, ritter_quantics_2024}. An alternative is to decompose $\Pi_{i,i+1}$ with a truncated SVD to get back to the MPS form with an adaptive bond dimension $\chi_i$, and then cross-interpolate the left and right site tensors. One sweeps through the whole system with such two-site updates until convergence of all local errors $\varepsilon_{i,i+1}=||\Pi_{i,i+1}-\tilde{\Pi}_{i,i+1}||$. Due to resemblance to the class of ground-state search algorithm for MPS, especially in the context of calculating an optimal, a priori unknown, local rank $\chi_i$ with SVD, the latter algorithm is often referred to as "DMRG" tensor cross interpolation. During the sweeps, one needs to keep track of the pivot indices already used to the left and right of the current $\Pi_{i,i+1}$ tensor to be able to link the global error of the whole tensor to the local error of the interpolation $\varepsilon_{i,i+1}$, which is referred to as the "nesting condition". In our subsequent calculations, we use the sweeping DMRG TCI algorithm variant with maxvol pivot search \cite{savostyanov_fast_2011} described as Algorithm 4 in \cite{dolgov_parallel_2020} and implemented in the \verb|torchTT| software package \cite{ion_torchtt_2025}. A full reconstruction of rank-$\chi$ MPS with TCI requires $\mathcal{O}(Nd^2\chi^2)$ function calls\footnote{We use the torchTT implementation which does not involve rook pivoting. Rook pivoting brings down the number of function calls to $\mathcal{O}(Nd \chi^2)$.}  \cite{dolgov_parallel_2020,fernandez_learning_2025}.

Although the DMRG cross interpolation algorithm proved to be effective in many cases, during sweeping the pivot matrix may become ill-conditioned, hence spoiling the convergence. Ref.~\cite{fernandez_learning_2025} introduces an attractive alternative to TCI, the partially rank-revealing LU decomposition, which is mathematically equivalent to TCI but numerically more stable since it avoids the pivot matrix inversion.

The open source community has been actively developing a number of software packages for TCI, with some notable examples including \verb|TT-Toolbox| (Matlab) \cite{TT-Toolbox}, \verb|ttcross| (Fortran, parallelized) \cite{ttcross}, \verb|ttpy| (Python) \cite{ttpy}, \verb|torchTT| (Python with torch \cite{paszke_pytorch_2019} backend) \cite{ion_torchtt_2025}, \verb|xfac| (C++ with Python bindings) \cite{xfac}, \verb|TensorCrossInterpolation.jl| \cite{TensorCrossInterpolation.jl} and \verb|ITensorNumericalAnalysis.jl| (Julia) \cite{tindall_itensornumericalanalysisjl_2025}.

\section{Gate optimization algorithms}
\label{app:gate_optimization}
\subsection{Evenbly-Vidal sweeping optimization algorithm}
\label{app:ev}

Let us consider a quantum circuit which realizes a state $U\ket{0}^{\otimes N}$ through a sequence of $M$ few-body unitary gates,
\begin{equation}
	U = U_M U_{M-1}...U_2 U_1.
\end{equation}
Our goal is to find gates $U_i$ that maximize the absolute value of the fidelity with a target state $\ket{\psi}$,
\begin{equation}
	F = \bra{000\dots 0}U^\dagger \ket{\psi},
\end{equation}
under constraint that the gates are unitary. Let us simplify the problem and concentrate on finding a single optimal gate $U_m$ while keeping the other gates fixed, with a method introduced in 2009 by Evenbly and Vidal \cite{evenbly_algorithms_2009} for isometric tensor networks, and later readapted by Shirakawa et al. for quantum circuits \cite{shirakawa_automatic_2024}. We can rewrite the fidelity as
\begin{equation}
	F_m = \bra{\phi_{m-1}} U_m^\dagger \ket{\psi_{m+1}},
\end{equation}
where
\begin{eqnarray}
	\ket{\psi_{m+1}} &=& U_{m+1}^\dagger U_{m+2}^\dagger \dots U_M^\dagger\ket{\psi},\\
	\bra{\phi_{m-1}} &=& \bra{000\dots 0} U_1^\dagger U_2^\dagger \dots U_{m-1}^\dagger.
\end{eqnarray}
Define the environment tensor for operator $U_m$,
\begin{equation}
	\mathcal{F}_m = \Tr_{\overline{U_m}} \left( \ket{\psi_{m+1}} \bra{\phi_{m-1}} \right)
	\label{eq:Fm2}
\end{equation}
where the trace goes over all sites on which the operator $U_m$ doesn't act. For example, for a two-site unitary $U_m$, Eq.~\eqref{eq:Fm2} gives an environment matrix $\mathcal{F}_m$ of size $4\times 4$. The $\mathcal{F}_m$ matrix is not unitary; in practice, it can be computed by removing the $U_m$ tensor from the tensor network representing the overlap \eqref{eq:F} and contracting the rest of the network, see Fig.~\ref{fig:evenblyvidal}a). If the state $\ket{\psi}$ is expressed in the MPS form, both $\ket{\psi_{m+1}}$ and $\bra{\phi_{m-1}}$ can be kept in the MPS form too, avoiding an exponential contraction cost of the entire network. Then, one can write an expression for the overlap as
\begin{equation}
	F_m = \Tr_{U_m}  \left( U_m^\dagger \mathcal{F}_m\right)
	\label{eq:Fm}
\end{equation}
where the trace goes around all sites on which $U_m$ acts. 

Let us perform the SVD of the environment tensor, 
\begin{eqnarray}
	\mathcal{F}_m = X D Y,
	\label{eq:fsvd}
\end{eqnarray}
where $X$, $Y$ are both unitary matrices and $D$ is a diagonal matrix of singular values.
It can be shown \cite{shirakawa_automatic_2024} that the optimal unitary $U_m$ maximizing the absolute value of fidelity in Eq.~\eqref{eq:Fm} is
\begin{equation}
	U_m' = X Y,
	\label{eq:evupdate}
\end{equation}
which is automatically unitary as $X$ and $Y$ are both unitary. Thus, the SVD of the gate environment tensor $\mathcal{F}$ gives a deterministic expression for an optimal gate $U_m'$ and requires an SVD decomposition of the environment matrix of the same dimension as the operator (in a typical case of a two-body operator, a $4\times 4$ matrix). Once the gate is updated, $U_m \gets U_m'$, one can proceed to the next gate $U_{m+1}$, which can be updated in a similar way. One does not need to recompute the entire environment from scratch, but reuse the intermediate states $\ket{\psi_{m+1}}$, $\bra{\phi_{m-1}}$ and "shift" them as 
\begin{eqnarray}
	\ket{\psi_{m+2}} \gets U_{m+1}\ket{\psi_{m+1}},\\
	\bra{\phi_{m}} \gets \bra{\phi_{m-1}} U_m'^\dagger,
\end{eqnarray}
while keeping their MPS form. An analogous operation can be performed if the move is done in the opposite direction, i.e. from $m$ to $m-1$. Visiting all $U_m$ twice, going from the first gate to the last and back, makes a single \textit{sweep}, which can be repeated $\nswpEV$ times until convergence.

It is important to note that even though a single step of the algorithm updates only a single local unitary, it takes into account all other gates as an environment of this unitary, thus the algorithm should be regarded as a global optimization method, especially after a complete sweep is done. 

\subsection{Riemannian variational optimization}
\label{app:riemannian}
\begin{figure}
	\centering
	\includegraphics[width=.5\linewidth]{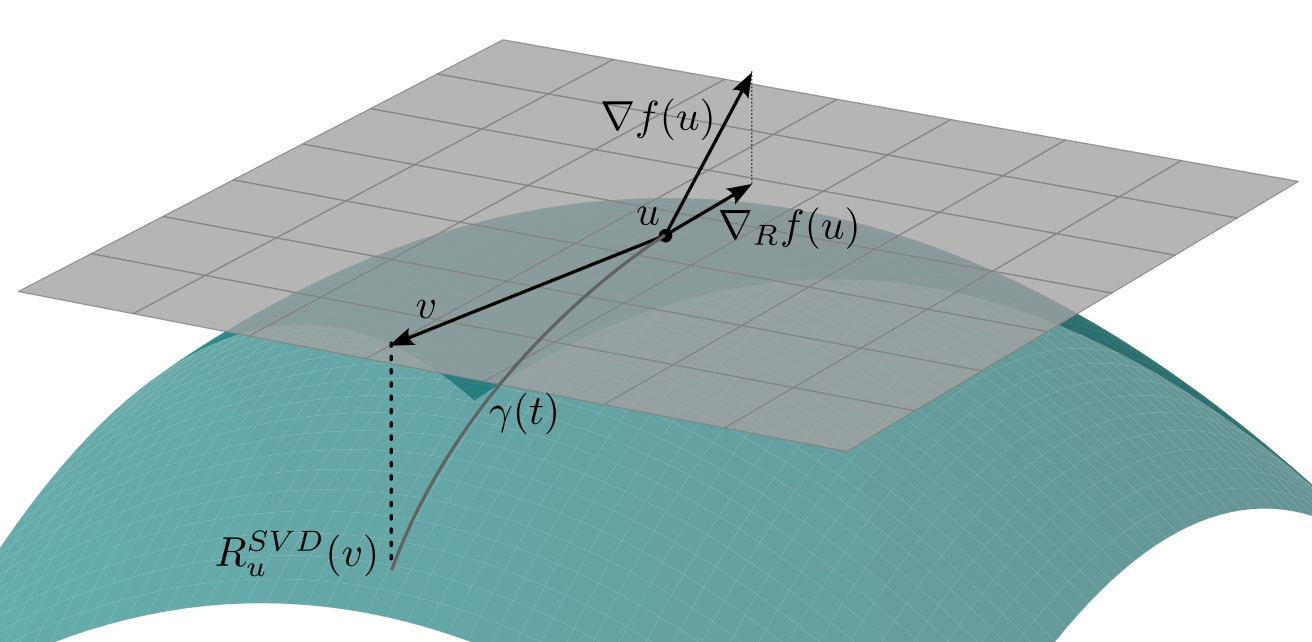}
	\caption{Riemannian variational optimization. The Euclidean gradient $\nabla f(u)$ does not lie in the tangent space at point $u$, unlike the Riemannian gradient $\nabla_Rf(u)$, see Eq.~\eqref{eq:riemanniangradient} for the Stiefel manifold. Transportation from point $u$ by a vector $v$, e.g. an update during optimization, is by definition along a geodesic $\gamma(t)$, $t\in[0,1]$. In order to avoid calculation of the geodesic, one takes its first order approximation $R^{SVD}_u(v)$ called retraction, which calculates the translated point by the  SVD of $u+v$, see Eq.~\eqref{eq:svdretraction}, up to an error $\mathcal{O}(||v||^2)$. For optimization on Riemannian manifolds one also introduces vector parallel transport, see text.}
	\label{fig:riemannian}
\end{figure}

In the following, we describe the necessary ingredients for Riemannian optimization, and give explicit formulas for the case of isometric matrices. Typically, during an optimization process, e.g. in the steepest descent method with momentum, one needs to be able to compute gradients of the objective at given points and transport these points (current solution) and vectors (e.g. current momentum) by a certain displacement vector (an update), all taking into account the local metric. In the Euclidean $\mathbb{R}^M$ space case, these operations are straightforward: we transport the point $u$ by vector $v$ as $u'=u+v$, the vectors do not change at all during the transportation to a different point, and the gradient of $f(u)$ at point $u$ is $\nabla f(u)$.

However, in the case of a manifold with a more complex structure, such as a  manifold of complex isometric matrices of size $n\times p$ (unitary matrices if $n=p$), called the Stiefel manifold,
\begin{equation}
	\text{St}_{n,p} = \left\lbrace V \in \mathbb{C}^{n\times p} | V^\dagger V = \mathbb{1} \right\rbrace,
	\label{eq:stiefel}
\end{equation}
the above operations need to take into account the local manifold geometry and the constraint satisfaction. The Euclidean gradient $\nabla f(u)$ needs to be substituted by the Riemannian gradient, which for the Stiefel manifold reads \cite{edelman_geometry_1998}
\begin{equation}
	\Delta_R f(u) = \frac{1}{2}u(u^\dagger \nabla f(u) - \nabla f(u) ^\dagger u) + \left(\mathbb{1} - u u^\dagger\right)\nabla f(u).
	\label{eq:riemanniangradient}
\end{equation}
The inner product of vectors $v,w$ at point $u$ is given by $\langle v,w\rangle_u =Re(\Tr(v^\dagger w))$.

The transport of a point $u$ along a vector $v$ would take us out of the manifold if we simply set $u'=u+v$ as in Euclidean case. Instead, the transport is defined in terms of a geodesic $\gamma(t)$ such that the final point is $u'=\gamma(1)$, with conditions $\gamma(0)=u$, $\frac{d\gamma(t)}{dt}_{t=0}=v$ \cite{absil_optimization_2008}. In practice, numerical calculation of the geodesic is typically inefficient \cite{zimmermann_computing_2022, sutti_shooting_2023}, and for small updates $v$ one uses the first-order appproximation of the geodesic called \textit{retraction},
$u' = R_u(v)$, see Fig.~\ref{fig:riemannian}. The retraction is not unique and a method can be chosen based on numerical efficiency. In the case of the Stiefel manifold, one can compute it using the SVD \cite{absil_optimization_2008},
\begin{equation}
	R^{SVD}_u(v) = XY,\quad \text{where }u+v = XDY,
	\label{eq:svdretraction}
\end{equation}
and $X,Y$ are isometries, $D$ a diagonal matrix of singular values. Note exactly the same form of an update from Eq.~\eqref{eq:evupdate} in the Evenbly-Vidal optimization algorithm. 

In case of the Stiefel manifold, the parallel transport of a vector, such as momentum in the momentum-based gradient descent algorithms, is performed by first transporting a point $u$ in which the vector is defined using retraction to get a new point $u'$, and then projecting the vector on the manifold in this new point  using the projection formula \cite{absil_optimization_2008, li_efficient_2020}
\begin{equation}
	P_{u'}(v) = \frac{1}{2}u'(u'^\dagger v - v^\dagger u') + (\mathbb{1}-u'u'^\dagger)v.
	\label{eq:projection}
\end{equation}
Notice that the Riemannian gradient in Eq.~\eqref{eq:riemanniangradient} is a result of projecting the Euclidean gradient onto the Stiefel manifold at point $u$.

Once we have the Riemannian gradient, we can perform gradient descent optimization. As an optimizer, we utilize the Adaptive Moment Estimation (Adam) method \cite{kingma_adam_2017}, widely used in machine learning, with default parameters $\beta_1=0.9$, $\beta_2=0.999$, $\epsilon=10^{-8}$. The optimizer calculates an adaptive learning rate for each parameter individually, based on the exponential moving averages of the gradient's first and second moment representing the optimization history. Internally, the optimizer takes all the modifications described above into account to stay on the Stiefel manifold. The Euclidean gradient, computed through autodifferentiation of the tensor network representing the circuit, is substituted with a Riemannian gradient from Eq.~\eqref{eq:riemanniangradient}. When an update $v$ is proposed for a point $u$, the new point is transported via SVD retraction, $u'=R^{SVD}_u(v)$. Also, parallel transport is performed on the momentum vector as described above.

The object that we optimize is the same as in the Evenbly-Vidal optimizer case, the state preparation fidelity parametrized by unitary gates $U_1,\dots U_M$. This defines a loss function in terms of a tensor network created by contraction of the MPS and the circuit. Contraction of the tensor network defines a computational graph which allows for autodifferentiation to obtain gradients of variational parameters \cite{liao_differentiable_2019}. For tensor network contractions, which are exponentially costly in general, we utilize the \verb|cotengra| hyperoptimized contractions \cite{gray_hyper_2021} with the \verb|cmaes| optimization library \cite{nomura_cmaes_2024}, \verb|kahypar| hypergraph partitioner \cite{schlag_high_2023}, no tensor slicing, and a subtree reconfiguration step \cite{huang_classical_2020, cotengra_tree_2025}.

In this work, the Riemannian optimization of quantum circuits on the Stiefel manifold part is performed using the \verb|QGOpt| software package \cite{luchnikov_qgopt_2021}. For other powerful and publicly available implementations in Python, see \verb|Pymanopt| \cite{townsend_pymanopt_2016} and \verb|Geoopt| \cite{geoopt_kochurov_2020}.

\section{1-to-2-qubit isometry decomposition}
\label{app:isometry}
\begin{figure*}
	\centering
	\includegraphics[width=.65\linewidth]{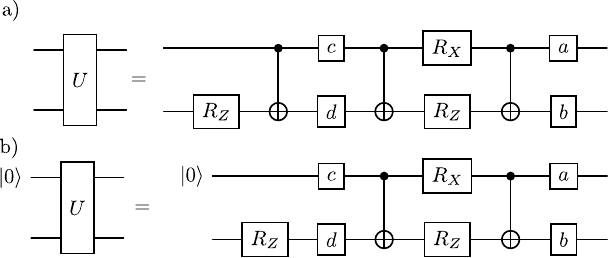}
	\caption{a) Decomposition of an arbitrary 2-qubit $SU(4)$ unitary gate \cite{shende_minimal_2004}. b) Decomposition of a 1-to-2-qubit isometry with the same construction as in a). Since the upper control qubit is in a definite initial state $\ket{0}$, the first CNOT gate on the left acts as an identity and can be removed, so one is left with 2 CNOT gates. $R_Z$ gate on the left can be absorbed into gate $d$.}
	\label{fig:gate_decomp}
\end{figure*}
Ref.~\cite{shende_minimal_2004} describes a decomposition of a 2-qubit SU(4) unitary into three CNOT gates, $R_{X,Z}$ rotations and single-qubit $a,b,c,d\in SU(2)$ gates,  as shown in Fig.~\ref{fig:gate_decomp}a). Proofs in Ref.~\cite{shende_minimal_2004} give an algorithm to compute $a,b,c,d$ and the three rotation angles. We use this construction to decompose 1-to-2 qubit isometries. By setting the upper qubit in Fig.~\ref{fig:gate_decomp}b) into a definite state $\ket{0}$, one can omit the first CNOT, since it would anyway act as an identity on the lower qubit. This leaves us with a construction requiring only 2 CNOT gates. The remaining $R_Z$ gate can be absorbed into the $d$ gate. Computation of all parameters is performed on the isometry with additional orthogonal columns which complete it to a unitary $U$. To obtain $U\in SU(4)$, $U$ is multiplied by a scalar, $U \rightarrow \det(U)^{-1/4} U \in SU(4)$. 

\section{SMPD convergence revisited}
\label{app:smpd_convergence}
\subsection{1D quantum Ising model groundstate at criticality}
\begin{figure*}
	\centering
	\includegraphics[width=\linewidth]{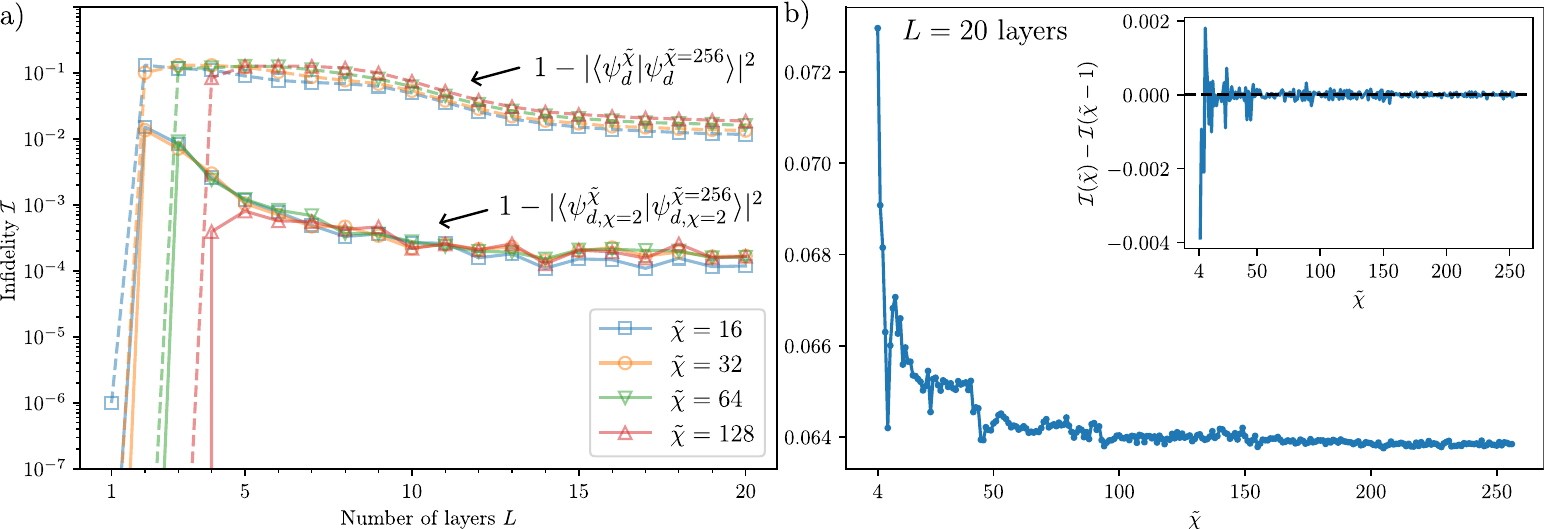}
\caption{Further study of the SMPD convergence. a) Infidelity of the disentangled state with its most accurate $\tilde{\chi}=256$ instance is quite substantial (upper curve), but its rank-2 approximation is orders of magnitude better (lower curve). The infidelities were computed using MPS simulator with bond dimension $\tilde{\chi}'=256$. b) For a fixed number of layers $L=20$, there are no gains from setting intermediate bond dimensions larger than $\tilde{\chi}\approx 50$. Inset shows differences of infidelities between consecutive values of $\tilde{\chi}$, stabilizing around zero.}
\label{fig:smpd_convergence}
\end{figure*}

Here we perform additional calculations with the critical quantum Ising model grounstate ($h_x=0.5$, $N=48$, $\chi=25$) as a state preparation target for SMPD, see Sec.~\ref{subsec:smpd_convergence}.

In Fig.~\ref{fig:smpd_convergence}a) we confirm that even though the accuracy of the rank-$\tilde{\chi}$ disentangled state $\ket{\psi_d^{\tilde{\chi}}}$ is quickly lost when increasing the number of layers $L$ (upper curve - we compute $\ket{\psi_d^{\tilde{\chi}}}$ infidelity with the one with maximal $\tilde{\chi}=256$), its rank-$2$ approximation $\ket{\psi_{d,\chi=2}^{\tilde{\chi}}}$ is robust to truncation errors (lower curve, infidelities $\Ical\sim10^{-4}$). To create the next layer of the disentangler, one uses this accurate rank-$2$ approximation, implying the gates are also accurately constructed even with low $\tilde{\chi}$.

In Fig.~\ref{fig:smpd_convergence}b) we concentrate on a relatively deep SMPD circuit by fixing the number of layers to $L=20$. We ask about the minimal $\tilde{\chi}$ which allows the preparation infidelity to reach the minimum and study the infidelity for varying ${\tilde{\chi}=4\dots 256}$. We observe that when $\tilde{\chi} \gtrapprox 50 = 2\chi$, the infidelity saturates, i.e., there is no benefit of constructing $L=20$ layer SMPD with higher intermediate bond dimensions $\tilde{\chi}$. The inset of Fig.~\ref{fig:smpd_convergence} illustrates that the differences of infidelity between neighboring $\tilde{\chi},~\tilde{\chi}-1$ stabilize around zero for $\tilde{\chi}\gtrapprox 50$. Interestingly, even an extremely low and numerically cheap $\tilde{\chi}=4$ gives a good approximation of the state with infidelity $\Ical\approx 0.073$. As a sanity check, we tested that the rank-4 approximation of the target state has an infidelity with the full state of $\Ical=0.0016$, which explains why the $\tilde{\chi}=4$ circuit approximation works so well.  

\begin{figure}[t!]
	\centering
	\includegraphics[width=.95\columnwidth]{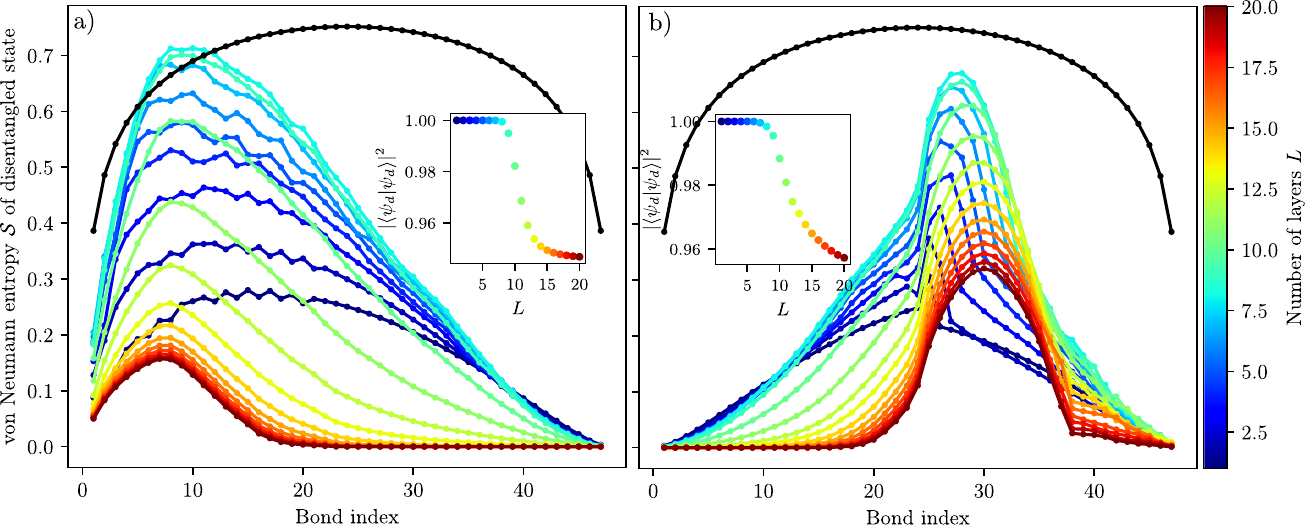}
	\caption{Evolution of the von Neumann bipartite entanglement entropy at each bond of the 1D transverse field Ising model groundstate ($h_x=0.5$, $N=48$, $\chi=25$, see Eq.~\eqref{eq:1dtfim} of the main text) disentangled by SMPD with a varying number of layers (colors). Black curve corresponds to the target state. Upon adding more layers, the entanglement on some bonds grows until $8$ layers (even exceeding the target state entropy around the $10$-th bond) and then starts to decrease. For an ideal disentangler, the entropy should reach $0$ everywhere. Bias towards low bond indices results from the right canonical form of the MPS. Final state preparation infidelity is $\Ical\approx 0.073$. We used maximal bond dimension $\tilde{\chi}=256$. Inset: norm of the truncated state after each disentangling layer. The state is normalized before the entropy calculation.}		
	\label{fig:entropies_disentangled_change}
\end{figure}
Inspired by Ref.~\cite{mansuroglu_preparation_2025}, we check how the entanglement entropy evolves during disentangling with SMPD circuits. In Fig.~\ref{fig:entropies_disentangled_change}a) we plot values of bipartite von Neumann entanglement entropy on all MPS bonds. The first layer decreases the entanglement roughly threefold. Then, adding new disentangling layers surprisingly leads to more entangled states, until $L=8$ layers are reached, and the state becomes guided towards the product state again. For a perfect disentangler, the entropy should reach $0$ at all bonds. Although the initial state has a symmetry in entropy with respect to the middle bond, the disentangled state loses this symmetry due to the right canonical form of the MPS. In the SMPD resulting from the right canonical form, one first applies disentangling gates starting from qubits $(i,i+1)=(47,48)$, then $(i,i+1)=(46,47)$, etc., down to qubits $(i,i+1)=(0,1)$, which seems to remove entanglement faster from bonds with a higher index. A completely symmetric behavior is observed for SMPD constructed from the left canonical form of the MPS, where the bonds with low indices are disentangled more efficiently.

Fig.~\ref{fig:entropies_disentangled_change}b) shows the same entanglement evolution during disentangling but for SMPD created from a mixed canonical form with orthogonality center placed at site $i=25$. It is clear that this affects the entanglement structure, now with more entanglement contained farther from the edges. Nevertheless, similarly to Fig.~\ref{fig:entropies_disentangled_change}a) the maximal entanglement grows until the $8$-th layer is reached, but it never exceeds the target state's entanglement. Afterwards, the entropy decreases towards zero.

The intermediate growth of the entanglement entropy is not problematic from the classical computation perspective - what matters for the construction of the next disentangler layer is a faithful rank-2 approximation of the disentangled state, not its full-rank form. This is analyzed in Sec.~\ref{subsec:smpd_convergence} of the main text.

\subsection{Random MPS}
\label{app:smpd_random_mps}
\begin{figure}
	\centering
	\includegraphics[width=.5\columnwidth]{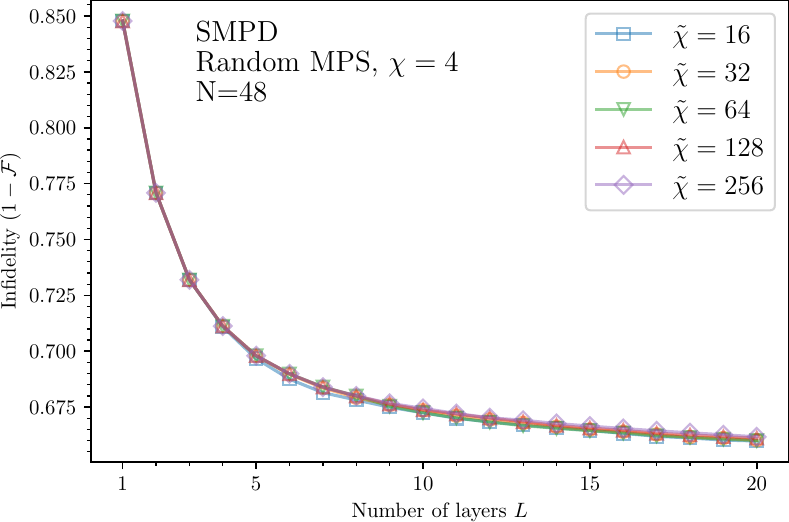}
	\caption{Infidelity as a function of the number of layers $L$ of the SMPD state preparation circuit for a complex-valued random MPS ($\chi=4$) target state on $N=48$ qubits. Although the infidelities reached by the SMPD algorithm are pretty large, the bond dimension $\tilde{\chi}$ required for the SMPD construction is much smaller than $2^L$ and there is no abrupt increse of infidelity beyond $L\gtrapprox\log_2\tilde{\chi}$. Infidelities are computed with bond dimension $\tilde{\chi}'=256$.}
	\label{fig:smpd_random_mps}
\end{figure}

We test our claims about SMPD circuit simulability from Sec.~\ref{subsec:smpd_convergence} of the main text, this time on a complex-valued random MPS target state with bond dimension $\chi=4$. We ask whether a much higher state preparation infidelity of the state preparation does not lead to problems with too large truncation errors for finite $\tilde{\chi}$. Results are presented in Fig.~\ref{fig:smpd_random_mps}. Indeed, we confirm that the SMPD disentangler with many layers is also simulable even for hard-to-prepare states such as random MPS. The infidelities were computed with bond dimension $\tilde{\chi}'=256$; the least accurate simulation in Fig.~\ref{fig:smpd_random_mps} has a norm error (i.e., accumulated truncation error in the state preparation) of $1.3\times10^{-5}$. 

\section{SMPD in the mixed gauge: orthogonality center position}
\label{app:mixed_pos}

\begin{figure*}[t!]
	\includegraphics[width=\linewidth]{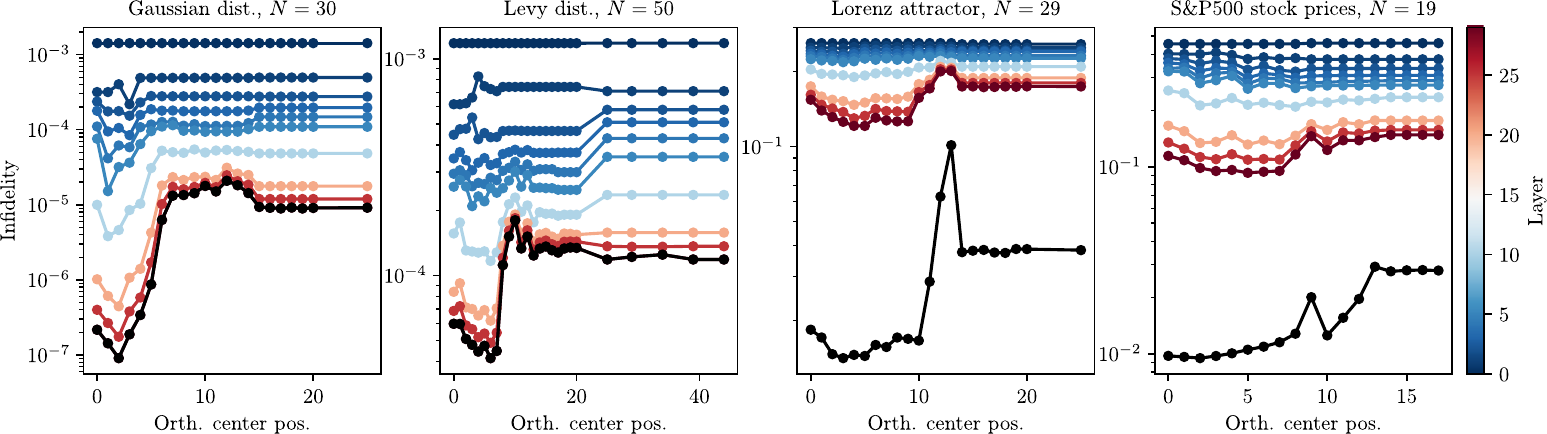}
	\caption{Infidelity as a function of the orthogonality center position for the sequential matrix product disentangler constructed upon the mixed gauge MPS representation. For all datasets, even though for the first few layers the position of the orthogonality center does not influence the infidelity, it is important to choose it optimally for deeper circuits, where one can gain even two orders of magnitude in the infidelity, see panel a). Black colors represent the last layer, i.e. 30th layer for Gaussian and Levy distributions, and 300th layer for the Lorenz attractor and S\&P500 datasets.}
	\label{fig:app:mixed_pos}
\end{figure*}

In Fig.~\ref{fig:app:mixed_pos} we stude the influence of the orthogonality center position on the infidelities reached with SMPDs. We confirm the finding of \cite{bohun_scalable_2025} that a correct placement of the orthogonality center can result in gains in terms of final infidelity. This is the case for the Gaussian distribution, for which we identify site $2$ as the optimal orthogonality center (this comes at a cost of deeper circuits as we are not in the center). For other datasets, the gains were not so significant, so we prioritize shallower circuits and place the orthogonality center in the middle of the system.

    \begin{figure}[t!]
	\centering
	\includegraphics[width=.77\columnwidth]{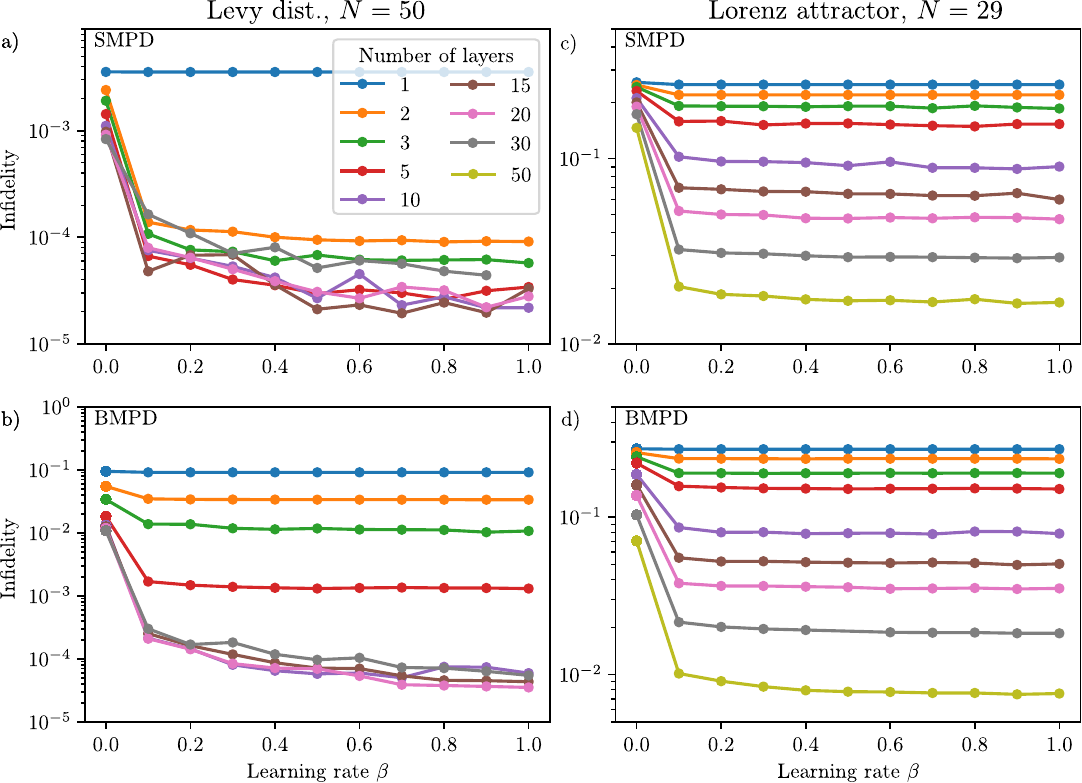}
	\caption{Final infidelity as a function of learning rate $\beta$ in the Evenbly-Vidal optimization ($\nswpEV=1000$ sweeps). a),c) SMPD-EV (mixed gauge, no gate decomposition); b),d) BMPD-EV circuits, with the L\'evy distribution (a), b)) and Lorenz attractor (c),d)) targets from Fig.~\ref{fig:datasets}. $\beta=0$ corresponds to no optimization. The impact of the choice of the learning rate on the final result is not substantial in both cases. For the calculations in main text, we choose $\beta=0.6$.  We use $\tilde{\chi}=32$ in a), b) and $\tilde{\chi}=128$ in c), d).
	}
	\label{fig:inf_vs_lr}
\end{figure}

\section{Evenbly-Vidal sweeping optimization: learning rate}   
    \label{app:learningrate}
	In Fig.~\ref{fig:inf_vs_lr} we test the influence of the learning rate $\beta$ from Eq.~\eqref{eq:learning_rate} in the EV optimization in two cases: the L\'evy distribution (a,b) and the Lorenz attractor (c,d), on both SMPD (a,c), and BMPD (b,d). The advantage of the global optimization starts to appear for $L>1$ layers - for single-layer disentangler circuits, the SMPD is optimal, and BMPD is almost optimal in terms of infidelity. In general, we find that for these states, the influence of $\beta$ on the final infidelity, in the range $\beta\in[0.2,1]$, is relatively weak. For the main text, we choose $\beta=0.6$.

\end{document}